\DocumentMetadata{}  

\PassOptionsToPackage{dvipsnames,table}{xcolor}
\PassOptionsToPackage{breaklinks=true,colorlinks=true}{hyperref}
\PassOptionsToPackage{nameinlink}{cleveref}

\documentclass[manuscript]{acmart}
\usepackage[british]{babel}
\usepackage[T1]{fontenc}
\usepackage[utf8x]{inputenc}

\usepackage{libertinus}

\RequirePackage{amsmath,amsfonts}
\usepackage{calrsfs}
\DeclareMathAlphabet{\pazocal}{OMS}{zplm}{m}{n}
\usepackage{subcaption,graphicx,wrapfig}
\usepackage{xspace,xifthen}
\usepackage[inline]{enumitem}
\usepackage{booktabs,multirow,array}
\usepackage[normalem]{ulem}
\usepackage[symbol*,hang,marginal]{footmisc}  
\usepackage{tikz,tikzsymbols,pgfkeys,adjustbox}
\usepackage{relsize}   
\usepackage{ragged2e}  
\usepackage{breakcites}
\usepackage{csquotes}
\usepackage{marginnote,tokcycle}

\RequirePackage{hyperref}
\RequirePackage{cleveref}

\AtBeginDocument{%
  \providecommand\BibTeX{{%
    \normalfont B\kern-0.5em{\scshape i\kern-0.25em b}\kern-0.8em\TeX}}}

\graphicspath{{figures/},}  

\newboolean{tosubmit}
\setboolean{tosubmit}{true}

\newboolean{showlinums}
\setboolean{showlinums}{false}

\ifthenelse{\boolean{showlinums}}{%
  \usepackage[pagewise]{lineno}  
  \linenumbers
  
}{}

\ifthenelse{\boolean{tosubmit}}{}{ 
  \usepackage[firstpage]{draftwatermark}  
  \SetWatermarkColor{Blue!11}
}

\Crefname{equation}{eq.}{eqs.}
\crefname{equation}{equation}{equations}
\Crefname{figure}{Fig.}{Figs.}
\crefname{figure}{figure}{figures}
\Crefname{tabular}{Table}{Tables}
\crefname{tabular}{table}{tables}
\Crefname{definition}{Def.}{Defs.}
\crefname{definition}{definition}{definitions}
\Crefname{proposition}{Prop.}{Props.}
\crefname{proposition}{proposition}{propositions}
\Crefname{section}{Sec.}{Secs.}
\crefname{section}{section}{sections}

\setquotestyle{latin}
\let\oldmktextquote\mktextquote
\renewcommand*{\mktextquote}[6]{\oldmktextquote{#1}{\textit{#2}}{#3}{#4}{#5}{#6}}

\DefineFNsymbols*{lamport}[math]{
  {\hspace{2pt}\reflectbox{\bfseries\textcopyright}\hspace{1pt}}
  {\hspace{1pt}\dagger\hspace{.2pt}}
  {\hspace{1pt}\ddagger\hspace{.2pt}}
  {\hspace{1pt}\ast\hspace{.2pt}}
  {\hspace{1pt}\S\hspace{.2pt}}
  {\hspace{1pt}\P\hspace{.2pt}}
  {\hspace{1pt}\|\hspace{.2pt}}
  {\hspace{1pt}\dagger\dagger\hspace{.2pt}}
  {\hspace{1pt}\ddagger\ddagger\hspace{.2pt}}
  {\hspace{1pt}\ast\ast\hspace{.2pt}}
}
\setfnsymbol{lamport}

\makeatletter
\def\THICKhrulefill{\leavevmode \leaders \hrule height 5pt\hfill \kern \z@}
\def\getfirst#1#2\relax{\tctestifnum{\count@stringtoks{#1}>1}{ERROR}{#1}}
\makeatother
\newcommand{\colorpar}[3]{\colorbox{#1}{\parbox{#2}{#3}}}
\newcommand{\marginremark}[3]{%
  \ifthenelse{\boolean{tosubmit}}{}{
	\marginnote{\raggedrightmarginnote\colorpar{#2}{.8\linewidth}%
      {\raggedrightmarginnote\color{#1}#3}}
}}
\newcommand{\textremark}[5]{%
  \ifthenelse{\boolean{tosubmit}}{}{
  \marginremark{#1}{#2}{\tiny\sffamily{[#3]\ #5}}%
  {\def\ULthickness{.8pt}\color{#1!80!black}\uline{#4}}
}}
\newcommand{\highlightedremark}[4]{%
  \ifthenelse{\boolean{tosubmit}}{}{
	\begin{center}\fcolorbox{#1}{#2}{%
	\begin{minipage}{.98\linewidth}\color{#1}%
	\textbf{\THICKhrulefill[ #3 ]\THICKhrulefill}%
	\par\noindent#4\end{minipage}}\end{center}%
}}
\newcommand{\hey}[4]{%
  \ifthenelse{\boolean{tosubmit}}{}{
  \reversemarginpar
  \leavevmode\marginnote{\sffamily\Large\color{#1}@\getfirst#3\relax\relax}
  \colorbox{#2}{\sffamily\bfseries{@#3:}}~{\sffamily\color{#1}#4}}
  \normalmarginpar}
\newcommand{\todo}[1]{%
  \ifthenelse{\boolean{tosubmit}}{}{
  \noindent\textsf{\color{Red}\textbf{TODO:} #1}%
  \marginnote{\textsf{\color{red}\bfseries TODO}}}}
\newcommand{\tocite}[1][??]{%
  \ifthenelse{\boolean{tosubmit}}{}{
  \noindent\textbf{\sffamily\textcolor{blue!85}{[#1]}}%
  \marginnote{\textsf{\color{blue}\bfseries CITE!}}}}
\colorlet{FM-fg}{Plum}
\colorlet{FM-bg}{Tan!12}
\colorlet{IP-fg}{BrickRed}
\colorlet{IP-bg}{orange!11}
\colorlet{CEB-fg}{TealBlue!75!green!75!black}
\colorlet{CEB-bg}{Aquamarine!8}
\colorlet{NNN-fg}{WildStrawberry!75!black}
\colorlet{NNN-bg}{Peach!33}

\newcommand{\heyCEB}[1]{\hey{CEB-fg}{CEB-fg!33}{Carlos}{#1}}

\newcommand{\hrmkCEB}[1]{\highlightedremark{CEB-fg}{CEB-bg}{CEB}{#1}}

\colorlet{shitbrown}{Brown!30!Maroon!90!black!90}
\NewDocumentEnvironment{shit}{ O{Fabio: ignore this!} }
  {\bigskip\begingroup\color{shitbrown}\slshape\smaller[1]}
  {\endgroup\medskip}

\def\exampletext{Example}
\newcounter{examplectr}
\newtheoremstyle{examplethmstyle}{1ex}{1ex}{\LibertinusSerifLF}{1em}{\LibertinusSerifSB\scshape\color{black!90}}{.}{.5em}{}
\theoremstyle{examplethmstyle}  
\newtheorem{example}[examplectr]{\expandafter\exampletext}
\theoremstyle{plain}        
\crefname{example}{example}{examples}

\usetikzlibrary{positioning,fit,patterns,patterns.meta}

\colorlet{hlboxcoldraw}{black!65}  
\colorlet{hlboxcolfill}{black!6}   
\newsavebox\BODYBOX  
\def\BODY{\unhbox\BODYBOX}
\NewDocumentEnvironment{hlbox}{ O{} }
  {\vspace{1.3ex plus .5ex minus .3ex}%
	\begingroup\centering\begin{spacing}{1.05}%
	\begin{lrbox}{\BODYBOX}\begin{minipage}{.97\linewidth}%
	\ifthenelse{\equal{#1}{}}{}{\textbf{#1}}}
  {\end{minipage}\end{lrbox}\begin{tikzpicture}%
	\node[rectangle,rounded corners=.3mm,inner sep=1.3ex,thick,
	      draw=hlboxcoldraw,fill=hlboxcolfill] {\BODY};
	\end{tikzpicture}\end{spacing}\endgroup%
	\vspace{1ex plus .5ex minus .3ex}}

\def\gaptext{Gap}
\newcounter{gapctr}
\newtheoremstyle{gapthmstyle}{.3ex}{.3ex}{\itshape}{0pt}{\bfseries}{.}{.5em}{\thmname{#1}\thmnumber{ #2}:\,{\thmnote{ #3}}}
\theoremstyle{gapthmstyle}  
\newtheorem{research-gap}[gapctr]{\expandafter\gaptext}
\theoremstyle{plain}        
\crefname{research-gap}{gap}{gaps}
\NewDocumentEnvironment{gap}{ O{You did not give this gap a title!} }
  {\begin{hlbox}\begin{research-gap}[#1]}
  {\end{research-gap}\end{hlbox}}

\NewDocumentEnvironment{gapalt}{ O{} }  
  {\gdef\title{#1}\vspace{1.5ex plus .5ex minus .3ex}
	\begin{research-gap}\begin{spacing}{0.95}\begin{lrbox}{\BODYBOX}}
  {\end{lrbox}\begin{center}\begin{tikzpicture}%
	[body/.style={rectangle,rounded corners=.7mm,semithick,
	              draw=hlboxcoldraw,fill=hlboxcolfill,inner sep=1.2ex,anchor=north west},
	 title/.style={rectangle,rounded corners=.3mm,semithick,
	               draw=hlboxcoldraw,fill=hlboxcolfill,anchor=north west},]
	\node [title] (TITLE) at (0,3.5ex) {%
		\bfseries \gaptext\:\thegapctr:~\title\vphantom{\big)}};
	\node [body] (BODY) at (0,0) {%
		\parbox[t]{.9\linewidth}{\raggedleft\BODY}};
	\node [rectangle,fill=hlboxcolfill,inner sep=0pt,
	       fit={([xshift=.6pt,yshift=3.55pt] TITLE.south west)
	            ([xshift=-.6pt,yshift=1.0pt] TITLE.south east)}] {};
	\end{tikzpicture}\end{center}\end{spacing}\end{research-gap}%
	\vspace{-2.2ex plus .5ex minus .3ex}}

\NewDocumentEnvironment{gapaltalt}{ O{} }
  {\gdef\title{#1}\vspace{1.5ex plus .5ex minus .3ex}
	\begin{research-gap}\begin{spacing}{0.95}\begin{lrbox}{\BODYBOX}}
  {\end{lrbox}\begin{tikzpicture}%
	[body/.style={rectangle,rounded corners=.7mm,inner sep=1.2ex,semithick,
	              draw=hlboxcoldraw,fill=hlboxcolfill,anchor=north west},
	 title/.style={rectangle,rounded corners=.3mm,semithick,
	               draw=hlboxcoldraw,fill=hlboxcolfill,anchor=north west}]
	\node [title] (TITLE) at (-.075\linewidth,4.2ex) {%
		\bfseries \gaptext\:\thegapctr:~\title\vphantom{\big)}};
	\node [body] (BODY) at (0,0) {%
		\parbox[t]{.9\linewidth}{\raggedleft\BODY}};
	\end{tikzpicture}\end{spacing}\end{research-gap}%
	\vspace{-2.2ex plus .5ex minus .3ex}}

\newcolumntype{L}[1]{>{\RaggedRight\let\newline\\\arraybackslash}m{#1}}
\newcolumntype{C}[1]{>{\Centering\let\newline\\\arraybackslash}m{#1}}
\newcolumntype{R}[1]{>{\RaggedLeft\let\newline\\\arraybackslash}m{#1}}

\colorlet{shade1}{black!9}
\colorlet{shade2}{white}
\def\colortablepreamble{%
  \renewcommand{\arraystretch}{1.1}
  \setlength{\tabcolsep}{3.5pt}
  \setlength{\aboverulesep}{-.25pt}
  \setlength{\belowrulesep}{.35pt}
  \setlength{\extrarowheight}{.3ex}
  \rowcolors{4}{shade1}{shade2}
}

\newlist{RQs}{enumerate}{1}
\setlist[RQs]{
	topsep     = .7ex,
	parsep     = .2ex,
	itemsep    = .5ex,
	leftmargin = 3.2em,
	label      = {\textsmaller[1]{\textbf{RQ\arabic*}}},
}
\Crefname{RQsi}{}{}

\newlist{objectives}{enumerate}{1}
\setlist[objectives]{
	topsep     = .7ex,
	parsep     = .2ex,
	itemsep    = 0pt,
	leftmargin = 2.6em,
	label      = {\textsmaller[1]{\bfseries(O\arabic*)}},
}
\Crefname{objectivesi}{Objective}{Objectives}

\newlist{assumptions}{enumerate}{1}
\setlist[assumptions]{
	topsep     = .5ex,
	parsep     = .1ex,
	itemsep    = 0pt,
	leftmargin = 3.3em,
	label      = \textsmaller[1]{\textsf{\bfseries A\arabic*~\:-}},
	ref        = \textscale{.9}{\textsf{A\arabic*}},
}
\Crefname{assumptionsi}{Assumption}{Assumptions}

\newlist{steps}{enumerate}{1}
\setlist[steps]{
	topsep     = .5ex,
	parsep     = .1ex,
	itemsep    = 0pt,
	leftmargin = 1.7em,
	label      = (\textscale{.93}{S}\arabic*),
	ref        = (\textscale{.93}{S}\arabic*),
}
\Crefname{stepsi}{Step}{Steps}

\newlist{exampleslist}{itemize}{1}
\setlist[exampleslist]{
	topsep     = 0pt,
	leftmargin = 1.2em,
	label      = {\raisebox{.5pt}{$\boldsymbol{\ast}$}},
	before     = \slshape,
}

\let\emptyset\varnothing
\newcommand{\from}{\colon}
\newcommand{\RR}{\ensuremath{\mathbb{R}}\xspace}  
\newcommand{\PDF}{\acronym{pdf}}             
\newcommand{\PDFs}{\PDF{s}\xspace}
\newcommand{\CDF}{\acronym{cdf}}             
\newcommand{\CDFs}{\CDF{s}\xspace}
\newcommand{\SF}{\acronym{sf}}               

\newcommand{\bfred}[1]{\textsf{\bfseries\color{Red}#1}\xspace}
\newcommand{\textsbf}[1]{{\libertineSB{#1}}\xspace}  
\newcommand{\code}[1]{\ensuremath{\text{\relscale{1.2}\ensuremath{\mathsmaller{\operatorname{\mathtt{#1}}}}}}\xspace}
\newcommand{\acronym}[1]{\ensuremath{\textsc{\larger{#1}}}\xspace}
\newcommand{\card}[1]{\ensuremath{\left\vert{#1}\right\vert}\xspace}
\newcommand{\tuple}[1]{\ensuremath{\left\langle{#1}\right\rangle}\xspace}

\newcommand{\Cpp}{C\raisebox{.3pt}{\kern-.4pt+\kern-.8pt+}\xspace}
\newcommand{\VCS}{\acronym{vcs}}             
\newcommand{\NVD}{\acronym{nvd}}             
\newcommand{\CVE}{\acronym{cve}}             
\newcommand{\CVEs}{\CVE{s}\xspace}
\newcommand{\EPSS}{\acronym{epss}}           
\newcommand{\FOSS}{\acronym{foss}}           
\newcommand{\ML}{\acronym{ml}}               
\newcommand{\AT}{\acronym{at}}               
\newcommand{\ATs}{\acronym{at}{s}\xspace}
\newcommand{\BAS}{\acronym{bas}}             
\newcommand{\AND}{\acronym{and}}             
\newcommand{\OR}{\acronym{or}}               
\newcommand{\DAG}{\acronym{dag}}             

\newcommand{\BDD}{\acronym{bdd}}             

\newcommand\restr[2]{{
  \left.\kern-\nulldelimiterspace 
  #1 
  \vphantom{\big|} 
  \right|_{#2} 
  }}
\DeclareMathOperator{\trelo}{\mathit{t}_\mathrm{rel}}
\DeclareMathOperator{\tstarto}{\mathit{t}_\mathrm{start\phantom{d}\!\!\!}}
\DeclareMathOperator{\tendo}{\mathit{t}_\mathrm{end}}
\DeclareMathOperator{\Treeo}{\mathit{D}}

\DeclareMathOperator{\abstraction}{\alpha}
\DeclareMathOperator{\ground}{\gamma}
\newcommand{\trel}{\trelo}
\newcommand{\nlgt}{\mathbin{\not\lessgtr}}
\newcommand{\tstart}{\ensuremath{\tstarto}\xspace}
\newcommand{\tend}{\ensuremath{\tendo}\xspace}
\newcommand{\lib}[1][]{\ensuremath{\ell\ifthenelse{\equal{#1}{}}{}{_{#1}}}\xspace}
\newcommand{\bflib}[1][]{\ensuremath{\boldsymbol{\ell}\ifthenelse{\equal{#1}{}}{}{_{\mkern-2mu{#1}}}}\xspace}
\newcommand{\clib}[2][]{\ensuremath{\ell\ifthenelse{\equal{#2}{}}{}{\scalebox{.9}{$\mkern-2mu\mathit{#2}\mkern2mu$}}\ifthenelse{\equal{#1}{}}{}{_{#1}}}\xspace}
\newcommand{\libdummy}{\lib[\mathsmaller{\boldsymbol{\times}}]}
\newcommand{\thisAT}[1][]{\ensuremath{\scalebox{.9}{$\pazocal{T}$}\ifthenelse{\equal{#1}{}}{}{_{\mkern-5mu{#1}}}}\xspace}
\newcommand{\thatAT}[1][]{\ensuremath{\scalebox{.9}{$\widehat{\pazocal{T}}$}\ifthenelse{\equal{#1}{}}{}{_{\mkern-5mu{#1}}}}\xspace}

\newcommand{\libsize}[1][]{\ensuremath{\card{\mkern1mu\lib[{#1}]}}\xspace}
\newcommand{\depsize}[1][]{\ensuremath{\libsize[{#1}]{\mkern-3mu}_\mathit{dep}}\xspace}
\newcommand{\ownsize}[1][]{\ensuremath{\libsize[{#1}]{\mkern-3mu}_\mathit{\vphantom{dp}own}}\xspace}
\newcommand{\totalsize}[1][]{\libsize[{#1}]}
\newcommand{\ttg}{\ensuremath{\mathtt{g}}\xspace}
\newcommand{\tta}{\ensuremath{\mathtt{a}}\xspace}
\newcommand{\ttv}{\ensuremath{\mathtt{v}}\xspace}
\newcommand{\ga}{\ensuremath{\ttg{:}\tta}\xspace}
\newcommand{\gav}{\ensuremath{\ttg{:}\tta{:}\ttv}\xspace}
\ExplSyntaxOn        
\seq_new:N\CVElist
\seq_set_from_clist:Nn\CVElist{%
	\ensuremath{\kappa},%
	\ensuremath{\varrho},%
	\ensuremath{\vartheta},%
	\ensuremath{\omega},%
	\ensuremath{\varepsilon},%
}
\newcommand{\CVEn}[1][]{\seq_item:Nn\CVElist{#1}\xspace}
\ExplSyntaxOff
\newcommand{\DepTree}[2][]{\ensuremath{\Treeo\ifthenelse{\equal{#1}{}}{}{_{#1}}({#2})}\xspace}
\newcommand{\TimeDepTree}[2][]{\DepTree[#1]{#2}}
\newcommand{\TimeDepTreeCut}[3][]{\ensuremath{\restr{\TimeDepTree[#1]{#2}}{#3}}\xspace}

\newcommand{\TDT}{\acronym{tdt}}       
\newcommand{\TDTs}{\acronym{tdt}{s}\xspace}

\newcommand{\CVSS}{\acronym{cvss}}

\newcommand{\SPOF}{\acronym{sp}{o}\acronym{f}}
\long\protect

\def\TITLESHORT{Forecasting the risk of software choices}
\def\TITLELONG{A model to foretell security vulnerabilities from library dependencies and source code evolution}
\hypersetup{%
    pdftitle={\TITLESHORT: \TITLELONG},
    pdfauthor={Carlos E. Budde, Ranindya Paramitha, Fabio Massacci},
    pdfkeywords={Library vulnerability prediction, risk models, cybersecurity},
}

\begin{document}

\title{\TITLESHORT}
\subtitle{\TITLELONG}

\author{Carlos E.\ Budde}
\orcid{0000-0001-8807-1548}
\email{carlosesteban.budde@unitn.it}
\author{Ranindya~Paramitha}
\orcid{0000-0002-6682-4243}
\email{ranindya.paramitha@unitn.it}
\affiliation{%
  \institution{University of Trento}
  \city{Trento}
  \country{Italy}
}


\author{Fabio Massacci}
\orcid{0000-0002-1091-8486}
%
\affiliation{%
  \institution{University of Trento}
  \city{Trento}
  \country{Italy}
}
\affiliation{%
  \institution{Vr\ij{e} Universiteit}
  \city{Amsterdam}
  \country{The Netherlands}
}

\renewcommand{\shortauthors}{Budde, Paramitha, Massacci}

\begin{abstract}
Software security mainly studies vulnerability detection: is my code vulnerable today?
This hinders risk estimation, so new approaches are emerging to forecast the occurrence of future vulnerabilities.
While useful, these approaches are coarse-grained and hard to employ for project-specific technical decisions.
We introduce a model capable of vulnerability forecasting at library level.
Formalising source-code evolution in time together with library dependency, our model can estimate the probability that a software project faces a CVE disclosure in a future time window.
Our approach is white-box and lightweight, which we demonstrate via experiments involving 1255 CVEs and 768 Java libraries, made public as an open-source artifact.
Besides probabilities estimation, e.g.\ to plan software updates, this formal model can be used to detect security-sensitive points in a project, or measure the health of a development ecosystem.
\end{abstract}

\ccsdesc[500]{Security and privacy~Formal security models}
\ccsdesc[500]{Security and privacy~Vulnerability management}
\ccsdesc[300]{Security and privacy~Software security engineering}
\ccsdesc[300]{Software and its engineering~Software evolution}


\maketitle

\setcounter{footnote}{1}  


\section{Introduction}
\label{sec:intro}

Avoiding or patching \emph{security vulnerabilities}---fragments of code vulnerable to exploits by a malicious party---is a prime concern in cybersecurity \cite{PSS+21,MW10,CKDR21}.
Reporting them, too: on Jan 2024, the European Commission adopted Regulation EU 2024/482, which defines the application of 
the \acronym{eucc}
as per the 2019 Cybersecurity Act~\cite{CyberSecAct,EUCC}.
Now, \acronym{eucc} certificate holders must report any vulnerability affecting their \acronym{ict} product to their national body, within 90 days of becoming aware of it.
And this picture is not unique to the EU: the legal duties of \acronym{ict} providers have been increasingly regulated in countries like Korea, the US, Norway, Israel, Singapore, etc.
In 2023, national bodies including the \acronym{nsa}, \acronym{bsi}, and \acronym{kisa}, co-authored \acronym{cisa}'s white paper \emph{Secure by Design}, 
which urges software manufacturers to
\textquote[SecureByDesign]{%
use a tailored threat model during the product development stage to address all potential threats to a system [\ldots] from the preliminary stages of design and development, through customer deployment and maintenance
}.

\def\SE{\acronym{se}}
Software vendors approach these duties in many ways.
The traditional reactionary policy remediates vulnerabilities as these are discovered in the code of our project---or any of its dependencies.
But this neglects that
(a) one cannot always replace, let alone fix, third-party code, and
(b) the sheer volume of vulnerabilities reported e.g.\ via a \emph{common vulnerability exposure} (\CVE) quickly becomes unmanageable \cite{MP21a}.
This despite metrics like the \emph{common vulnerability scoring system}, which prioritises \CVEs by severity but struggles to tell whether they will lead to actual attacks \cite{SHH+21}.

A recent notable tool in this direction is the \emph{Exploit Prediction Scoring System} (\EPSS): \textquote{a completely data-driven} approach to \textquote[JRS+23]{distinguishing vulnerabilities that are exploited in the wild and thus may be prioritized for remediation}. 
The \EPSS \textquote{leverages 1,477 features for predicting exploitation} including \acronym{mitre} \CVEs, \acronym{cisa}'s \emph{Known Exploited Vulnerabilities} catalog (\acronym{kev}), talk of an exploit in social media, etc.~\cite{JRS+23}.
Yet it remains reactionary in nature because \emph{predict} is machine-learning jargon for \emph{detect}: to tell if something is already there, using e.g.\ hindsight data such as \acronym{kev}. 

Thus, the \EPSS---and any use of \ML to \textquote{predict} vulnerabilities---helps to guess whether a vulnerability that affects our code today, is being exploited today and can impact our project.
However, this does not address prevention as laid out e.g.\ in \emph{Secure by Design}:
hindsight can help to improve ``next time'', e.g.\ to design a new project, but prevention in current projects calls for anticipation of exploits before these occur---some oracle for vulnerabilities to come~\cite{Mas24}.

\begin{figure}
    \centering
    \includegraphics[width=.95\linewidth]{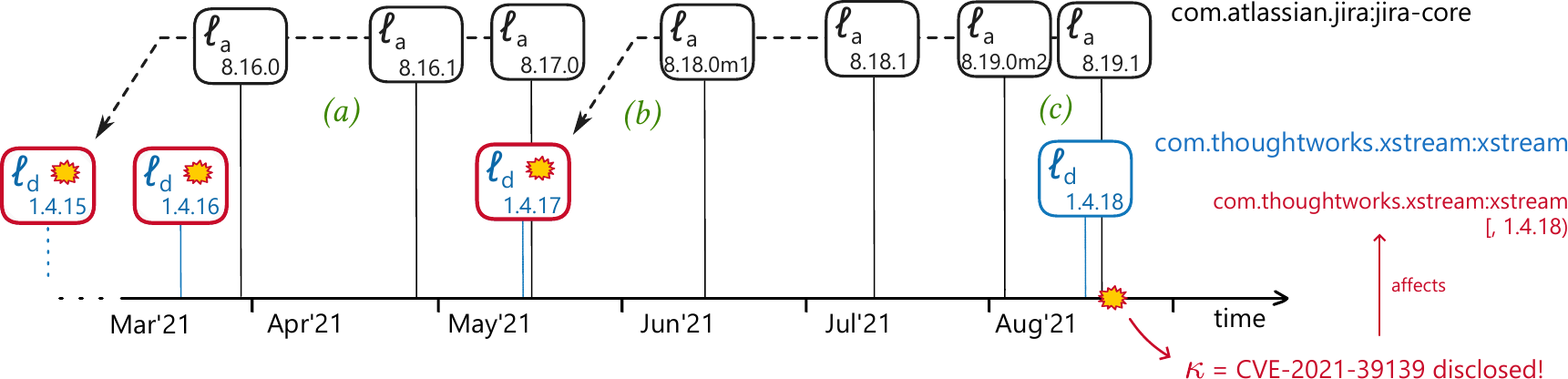}
	\begin{minipage}{.96\linewidth}
		\smaller[.5]
		Codebases like $\lib[a]=\cdots\code{jira-core}$ update their
		dependencies following company policies:
		\begin{enumerate}[topsep=1pt,itemsep=.4ex,leftmargin=2.5em,
		                  label=\textit{\color[RGB]{61,134,0}(\parbox{.5em}{\centering\alph*})}]
		\item	most updates of \lib[a] decide to keep the same (old)
				version of dependency \lib[d];
		\item	at certain points, the latest available version of \lib[d] is adopted;
		\item	\lib[a] is affected by \CVEn[1] via \lib[d]---this could have been
				avoided, but \CVEn[1] was unknown when update decisions were made.
				\label{toolate}
		\end{enumerate}
		\smallskip
		Estimating the probability of vulnerabilities like \CVEn[1]
		\emph{in the future}---i.e.\ before \ref{toolate}---is key
		to improve dependency-update policies.
	\end{minipage}
	\hspace{-1ex}
    \caption{Update policies that disregard the probability of future
		vulnerabilities threaten the security of entire software projects}%
    \label{fig:motivating_example}
	\vspace{-2ex}
\end{figure}

A few recent successes have shown this task to be feasible, e.g.\ estimating the number of \CVEs expected in coming years~\cite{LRW22,YPWS20}.
Such high-level metrics are however difficult to apply to specific codebases, e.g.\ to decide whether a dependency shall be updated or otherwise changed.
This is illustrated in \Cref{fig:motivating_example}, where the developers of ${\lib[a]=\code{com.attlassian.jira:jira-core}}$ would want to know, when planning the release of their next version, whether ${\lib[d]=\code{com.thoughtworks.xstream:xstream}}$ should be kept as is, or updated e.g.\ to its latest version.
Making the wrong decision can force developers to factor-in unplanned patches, as happened to \lib[a] in \Cref{fig:motivating_example} after time \textit{(c)}.

Such finer-grained decisions call for metrics akin to the ones used in software engineering to measure code quality.
But while cyclomatic complexity and code churn can be used to estimate the future cost of maintaining a codebase, such type of metrics have not been able to foretell vulnerabilities 
in general applications~\cite{LYZ+23,PSS+21,MW10,CKDR21}.

\begin{hlbox}[RQ:]
	Is it possible to measure the likelihood of discovering new (future) vulnerabilities in a specific project?
\end{hlbox}

Technically, any metric that answers positively this research question must avoid label creep: data used to foretell an unseen vulnerability \CVEn[1] in a future time window $T=\left[\tstart,\tend\right]$ must exclude information on the occurrence of \CVEn[1] during $T$. 
Thus, fitting/learning cannot use hindsight data such as \CVE or \acronym{kev} entries for \CVEn[1].
%
\emph{This work introduces an estimate for the probability of observing \CVEn[1], fitted on past data from comparable codebases of the entire dependency tree.}

\subsection*{Contributions}
\label{sec:intro:contributions}

\begin{itemize}[leftmargin=1.7em,topsep=1ex,itemsep=1ex,]
\item	\textbf{Estimators} for the probability of having a \CVE disclosed in
		the future for a given codebase, as probability density functions
		(\PDF) with support on future time, i.e.\ the semi-line $\mathbb{R}_{\geqslant0}$.
		\begin{itemize}[label=\raisebox{1pt}{$\mathsmaller\blacktriangleright$},parsep=.5ex,itemsep=.3ex]
		\item	This can be used in situations like \Cref{fig:motivating_example},
				providing quantities useful for company policies---e.g.\
				for the adoption of dependencies updates---to make
				informed decisions from the cybersecurity perspective.
		\item	One \PDF describes the likelihood of vulnerabilities
				in one codebase: we also present an analytical approach
				to extend our estimators for multi-codebase analysis.
		\end{itemize}
\item	\textbf{Time Dependency Trees} (\TDT): a white-box model that uses the
		above \PDFs to estimate potential vulnerabilities in the dependency
		tree of a software project, yet avoiding the full expansion of the
		probability space.
		\begin{itemize}[label=\raisebox{1pt}{$\mathsmaller\blacktriangleright$},parsep=.5ex,itemsep=.3ex]
		\item	\TDT-based estimations leverage \PDF metrics by adding a
				temporal component to the project-level analysis,
				and using it to estimate the probability that a
				full dependency tree faces a \CVE in the future.
				This is a lightweight alternative to the analytical approach
				from above, applicable to projects with dozens of dependencies.
		\item	\TDTs, as a minimal graph representation of dependencies
				across time and software codebases, allow for further
				project- and even ecosystem-level studies, that depend on the
				evolution in time of entire software codebases.
		\end{itemize}
\item	\textbf{A software artifact} for experimental reproduction of our work
		\cite{BPM24}, in compliance with ACM terminology \cite{ACM_artifact}.
\end{itemize}

Inspired by practical applicability, \TDTs revise the hypothesis that, in cross-project and cross-ecosystem analyses, code metrics can be meaningfully correlated to fine-grained vulnerability data such as type, location, and variables.
Like economics and sociology, which can foretell to some extent the behaviour of masses of people but not of individuals, our results posit that code metrics can be used for vulnerability analysis at high level, e.g.\ to estimate the number of vulnerabilities occurring in a characterised codebase in a time span.
Marking e.g.\ individual vulnerable code fragments falls out of scope, and is expected to remain practically infeasible for project-agnostic approaches.
			
Besides introducing \TDTs, this work uses them for a practical demonstration of \CVE forecasting.
Technically, this starts from the traditional dependency tree(s) of a project under study, which are then transformed into attack trees.
The leaves of the tree are decorated with precomputed \CDFs, instantiated in time points relatives to the release date of the codebases that each leaf represents.
This information is then aggregated into an estimate of the (joint) probability of the fastest expected \CVE disclosure affecting the entire project.

Nevertheless, vulnerability forecasting is one of the possible applications of \TDTs.
\Cref{sec:model} mentions how they can also be used to:
\begin{enumerate*}[label=(\alph*),itemjoin={{; }}]
\item	pinpoint the main suspects (libraries)
		of expected forthcoming vulnerabilities
\item	discover dangerous pervasive dependencies
		across several versions of a codebase
\item	assess the health of a development ecosystem;
\end{enumerate*}
etc.

\section{Research Questions and Objectives}
\label{sec:rq}

This \namecref{sec:rq} further elaborates the example from \Cref{fig:motivating_example}, refining our original research question into two subquestions and matching research objectives.
We begin by defining the terminology used in the rest of the article.

\subsection{Terminology}
\label{sec:rq:terminology}

We rely on the terminology established among practitioners---e.g.\ users of Apache Maven---and consolidated in~\cite{PPP+18,PPP+22}:

\begin{itemize}[leftmargin=1em]
%
\item	
\textbf{Libraries:}
A \emph{library}, denoted \lib, $\lib'$, \lib[x], $x$, $a$, etc.\ is a separately distributed software component, typically consisting of a logically grouped set of classes (or functions, or APIs, etc.) that may be provided in a single file, or a set of files.
\item	\textbf{Library instances:}
The term ``library'' on its own is version- and time-agnostic.
	A specific version of a library is a \emph{library instance}---we identify them via subindices that can be numeric or $i, j, k, v_i$, etc. E.g.\ \lib[1], \lib[a_{1.01}], \lib[v_i], $x_{v_1}$, etc.
	\begin{itemize}[leftmargin=1.5em]
	\item	In a set of library instances
			$\{\lib[a_i],\lib[a_j],\lib[a_k],\ldots\}$,
			subindices $i,j,k$ indicate the different versions
			of library \lib[a].
	\item	To simplify our presentation, for each library we assume a
			total order of versions (e.g.\ $\lib[a_i]<\lib[a_j]<\lib[a_k]$),
			where the order of a version is given by its release date s.t.\ 
			instances with a later release date have higher order.\!%
			\footnote{%
			Instances released on the same date can be disambiguated via
			\texttt{MAJOR.Minor.patch} version comparison \cite{SemVer}.
			This can also be easily generalised to the case of divergent
			versions---e.g.\ parallel releases of Apache Tomcat for
			development branches v8.5 and v9.0---via c-chains,
			as shown in \Cref{sec:model}.}
	\end{itemize}
\item	\textbf{Dependencies:}
A \emph{dependency} is a library instance whose functionalities are used by another library instance---e.g.\ if the execution of a method in \lib[a_1] uses a function from \lib[b_2], directly or indirectly, then \lib[b_2] is a dependency of \lib[a_1].
    \begin{itemize}[leftmargin=1.5em]
	\newlength{\strutheight}
	\settoheight{\strutheight}{\strut}
	\begin{adjustbox}{raise=\strutheight,minipage={\linewidth}}
      \begin{wrapfigure}[9]{r}{.51\linewidth}
		\raggedright\hspace{2pt}\vspace{-1ex}%
        \includegraphics[width=.93\linewidth]{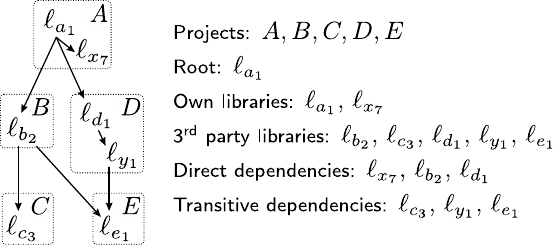}

		\vspace{-.7ex}%
		\caption{\textbf{Dependency tree}: arrow means ``uses in source code''}
		\label{fig:dependency_tree}
      \end{wrapfigure}
	\item	The \emph{dependent (library) instance} is the reciprocal \strut
			of the dependency, e.g.\ if \lib[b_2] is a dependency of \lib[a_1],
			then \lib[a_1] is a dependent instance of \lib[b_2].
			\begin{exampleslist}
			\item	In \Cref{fig:dependency_tree}, \lib[b_2], \lib[c_3], and
					\lib[x_7], are dependencies of \lib[a_1].
			\item	In turn, \lib[a_1] is a dependent instance of them.
			\end{exampleslist}
    \item   A \emph{direct dependency} is a dependency called directly in
			the source code of the dependent instance---e.g.\ \lib[b_2] is
			a direct dependency of \lib[a_1] if the latter invokes a function
			of the former in its source code.
			\begin{exampleslist}
			\item	In \Cref{fig:dependency_tree}, \lib[b_2] is a direct
					dependency of \lib[a_1], and \lib[c_3] is a direct
					dependency of \lib[b_2].
			\end{exampleslist}
	\item	A \emph{transitive dependency} is a dependency that is not direct%
			---e.g.\ if \lib[c_3] is not a direct dependency of \lib[a_1], but
			\lib[a_1] uses a function of \lib[b_2] that in turn uses a method
			from \lib[c_3], then \lib[c_3] is a transitive dependency of
			\lib[a_1].
			\begin{exampleslist}
			\item	In \Cref{fig:dependency_tree}, \lib[c_3] and \lib[e_1]
					both are transitive dependencies of \lib[a_1].
			\end{exampleslist}
	\end{adjustbox}
    \end{itemize}
\item	\textbf{Dependency tree:}
A \emph{dependency tree} ($T$) \strut is a connected, single-rooted, directed acyclic graph (\DAG) whose nodes are library instances, and whose edges point from each instance to its direct dependencies.
	\begin{itemize}[leftmargin=1.5em]
	\item	The \emph{root} or \emph{main library} in $T$ is
			its root node: the only library instance that is not
			a dependency of any other.
			\begin{exampleslist}
			\item	The root of the dependency tree in
					\Cref{fig:dependency_tree} is the library
					instance \lib[a_1], pictured on top as usual.
			\end{exampleslist}
	\item	The \emph{leaves} in $T$ are library instances that have no
			dependencies.
			These usually correspond to base libraries of the language,
			such as the Standard C++ Library, the Java or Python
			Standard Libraries, etc.
			\begin{exampleslist}
			\item	The leaves in \Cref{fig:dependency_tree} are the library
					instances \lib[x_7], \lib[c_3], and \lib[e_1].
			\end{exampleslist}
    \end{itemize}
\item	\textbf{Projects:}
A \emph{project} is a set of libraries developed and/or maintained together by a group of developers.
	\begin{itemize}[leftmargin=1.5em]
	\item	\emph{Own dependencies} for a project are those belonging to
			the same project as the dependent library instance.
			This can be lifted to dependency trees, where own dependencies
			are those belonging to the same project as the root.
	\item	\emph{Third-party dependencies} for a project are, instead,
			dependencies maintained by other projects.\!%
			\footnote{%
			Developers may structure their own code
			in separate libraries, which must not be counted as other
			project's code but as its own code \cite{PPP+22}.}
			\begin{exampleslist}
			\item	In \Cref{fig:dependency_tree}, project $A$ is formed by
					\lib[a_1] and \lib[x_7] alone, so \lib[x_7] is an own
					dependency while all the rest are third-party.
			\end{exampleslist}
	\end{itemize}
\item	\textbf{Code size:}
For a given library instance \lib[j]:
	\begin{itemize}[leftmargin=1.5em]
	\item	Its \emph{own code size}, \ownsize[j], is the summed number of
			lines of code from all files that form \lib[j] and its own
			dependencies.
	\item	Its \emph{dependency code size}, \depsize[j], is the sum of
			lines of code of third-party direct and transitive dependencies.
	\item	Its \emph{total code size} is
			$\totalsize[j] = \ownsize[j] + \depsize[j]$.
			\begin{exampleslist}
			\item	In \Cref{fig:dependency_tree}, \ownsize[a_1] are the lines
					of code of \lib[a_1] plus those of \lib[x_7],
					whereas \depsize[a_1] considers
					\lib[b_2], \lib[c_3], \lib[d_1], \lib[y_1], and \lib[e_1].
			\end{exampleslist}
	\end{itemize}
\item	\textbf{Vulnerabilities:}
A \emph{vulnerability} is an identification of undesired behaviour in a piece of software, that can be exploited by (i.e.\ caused and used to the malicious intent of) an attacker:
	\begin{itemize}[leftmargin=1.5em]
	\item	We study publicly disclosed vulnerabilities, such as \CVEs,
			which we refer to using Greek letters (\CVEn[1], \CVEn[2],
			\CVEn[3], \CVEn[4], \ldots) or their named IDs
			(e.g.\ CVE-2014-0160).
	\item	The \emph{vulnerable set} of vulnerability \CVEn[1] is the
			set $\Lambda_{\CVEn[1]}\neq\emptyset$ of library instances
			whose source code allows the exploit.
	\item	\emph{Vulnerability propagation} refers to the fact that, if
			$\lib[b_2]\in\Lambda_{\CVEn[1]}$ and \lib[b_2] is a dependency
			of \lib[a_1], then \lib[a_1] can be affected by an attacker that
			exploits \CVEn[1] even though $\lib[a_1]\notin\Lambda_{\CVEn[1]}$.
	\end{itemize}
\end{itemize}

\smallskip
\noindent
We also use standard probability theory as the basis for our vulnerability-forecasting model:

\begin{itemize}[leftmargin=1em]
\item	\textbf{Probabilities}:
A \emph{probability space} is a triple $(\Omega,\mathcal{B},\mu)$ where $\Omega$ is the sample space, $\mathcal{B}$ is the Borel $\sigma$-algebra on $\Omega$, and $\mu\from\mathcal{B}\to[0,1]$ is a probability function, i.e.\ 
	\begin{itemize*}[label=,itemjoin={{, }},itemjoin*={{, and}}]
	\item	$\mu(\emptyset)=0$
    \item   $\mu(\bigcup_{i\in I} A_i) = \sum_{i\in I}\mu(A_i)$ for any denumerable set $\{A_i\}_{i\in I}$ of disjoint events $A_i\in\mathcal{B}$
    \end{itemize*}.
\item	\textbf{PDF}, \textbf{CDF}, \textbf{SF}:
A \emph{probability density function} (\PDF) is a probability function $P$ whose sample space is the semiline~$\RR_{\geqslant0}$.
By the conditions above it follows that $P(\{x\})=0$ for any point $x\geqslant0$.
	\begin{itemize}[leftmargin=1.5em]
	\item	A \emph{cumulative density function}, $\CDF\from\RR_{\geqslant0}\to[0,1]$, is defined for every \PDF as: $\CDF(X) \doteq \int_{x=0}^{X} \PDF(x)\,dx$.
    \item   A \emph{survival function}, $\SF\from\RR_{\geqslant0}\to[0,1]$, is defined for every \PDF as: $\SF(X) \doteq 1-\CDF(X)$.
    \end{itemize}
\end{itemize}

\subsection{Motivating Example and Research Questions}
\label{sec:rq:motivating_example}

\begin{figure}
    \centering
    \includegraphics[width=.95\linewidth]{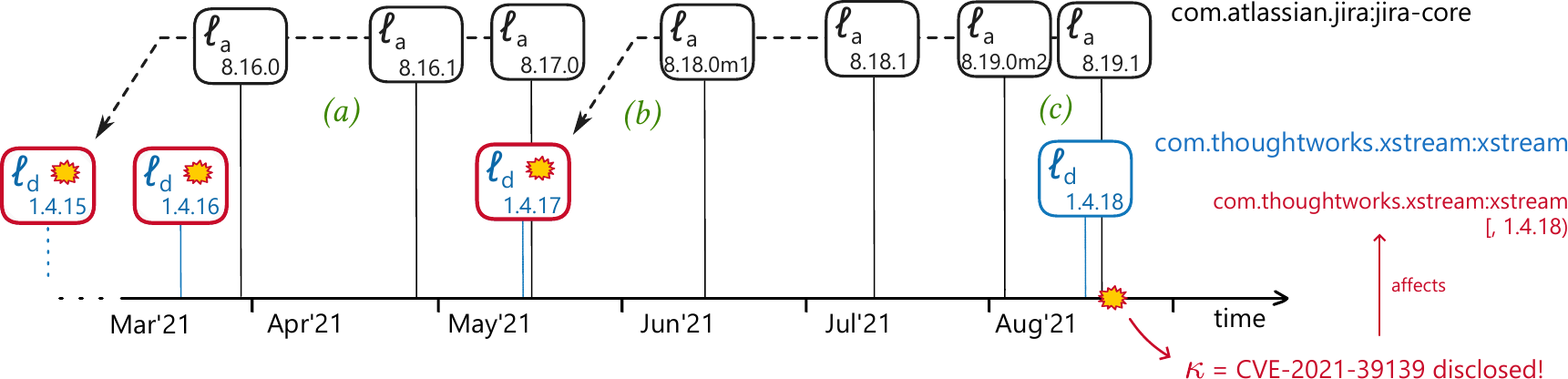}
    \caption{Update policies of \lib[a] need forecasting metrics to avoid
	         future vulnerabilities from dependencies like \lib[d]
			 (same as \Cref{fig:motivating_example})}
    \label{fig:motivating_example_copy}
\end{figure}

\begin{example}
\label{ex:motivating_example}
To motivate vulnerability forecasting we use the example of \Cref{fig:motivating_example_copy}.
Found in the wild, it concerns the Jira project which is one of the most popular bug tracker in the world.
The core artifact, $\lib[a]=\code{jira-core}$, uses the dependency ${\lib[d]=\code{com.thoughtworks.xstream:xstream}}$ for XML serialisation.
While the upstream version of \lib[a] is updated monthly, new versions of \lib[d] are adopted much less frequently.
In August 2021 two versions of \lib[a] were released but neither updated \lib[d], which was kept at version 1.4.17.
A vulnerability was then disclosed, $\CVEn[1]=\text{\sffamily CVE-2021-39139}$ with $\CVSS=8.8$, which affected all versions of \lib[d] except for the recently-released \code{xstream:1.4.18}.
The timeline in \Cref{fig:motivating_example_copy} shows how \lib[a] thus became vulnerable to the security bug \CVEn[1], which it had had the opportunity to avoid---by updating to the latest version of \lib[d]---but for which it had no hint of its forthcoming.
\end{example}

\begin{hlbox}[RQ1:]
	Can we forecast situations like that of \Cref{fig:motivating_example_copy},
	via metrics that quantify their likelihood in the future?
\end{hlbox}

This research question is addressed by the following objective:
\begin{objectives}
\item   \label{obj:metric}
        Quantify the probability that a software project faces
        a security vulnerability in the future:
        %
        \begin{itemize}[label=\textbullet,leftmargin=1.2em]
        \item   this metric should be parameterised by the (future)
                time horizon captured by the prediction;
        \item   it should take into account the vulnerabilities that
                would affect any dependency on the project.
        \end{itemize}
\end{objectives}

\begin{figure}
  \centering

  \begin{subfigure}[t]{.45\linewidth}
	\centering
	\includegraphics[width=.83\linewidth]{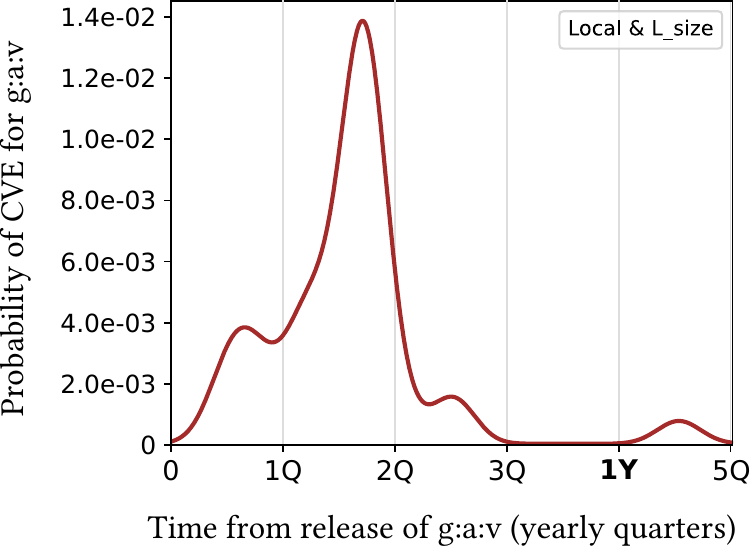}
	\caption{Probability of \CVE disclosure for \lib[d]
		from \Cref{fig:motivating_example_copy}}
	\label{fig:holy_grail:1Dplot}
  \end{subfigure}
  \qquad
  \begin{subfigure}[t]{.45\linewidth}
	\centering
	\includegraphics[width=.95\linewidth]{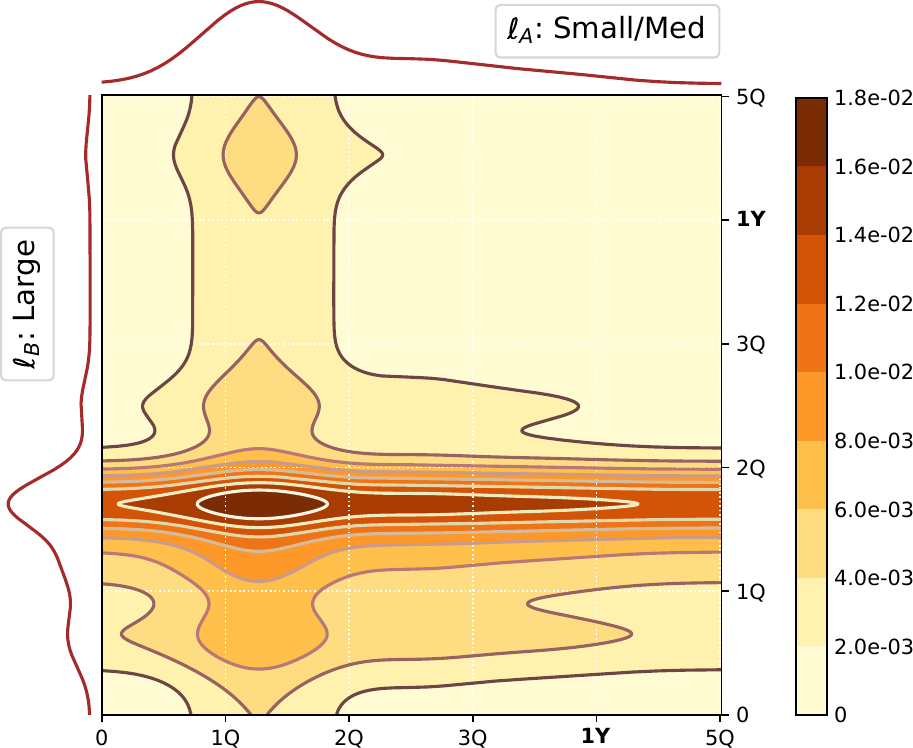}
	\caption{Contour plot for the joint probability of \CVE disclosure
		for two libraries at any time point}
	\label{fig:holy_grail:2Dplot}
  \end{subfigure}

  \begin{tikzpicture}[overlay]
	\draw[thick,color=black!40,densely dashed] (-0.98,5.07) -- (1.11,5.6);
	\draw[thick,color=black!40,densely dashed] (-0.98,1.71) -- (1.11,1.2);
	\definecolor{rojo}{HTML}{C80B2A}
	\definecolor{amarillo}{HTML}{FFCC00}
	\node (KAPPA-POINT) at (-4.45,2.55) [shape=circle,thick,draw=rojo,
		fill=amarillo,fill opacity=.3,inner sep=2pt] {};
	\node (KAPPA-TEXT) at (-3,5.6) {\textcolor{black}{\CVEn[1]}
		\smaller published for \lib[d] = \code{xstream:1.4.17}};
    \node (KAPPA-POINT-ON-PLANE) at (1.02,2.04) [shape=circle,draw=rojo,
		fill=amarillo,fill opacity=.3,inner sep=1.5pt] {};
	\draw[->,color=rojo,thick,dotted]
		([xshift=-2.5pt,yshift=2pt] KAPPA-POINT.north)
			.. controls +(-6pt,5pt) and +(-11pt,-20pt) ..
		([xshift=1.5pt,yshift=1pt] KAPPA-TEXT.south west);
	\draw[->,color=rojo,thick,dotted]
		([xshift=2pt,yshift=2pt] KAPPA-POINT.east)
			.. controls +(30pt,15pt) and +(-16pt,-5pt) ..
		([xshift=-2pt,yshift=-1pt] KAPPA-POINT-ON-PLANE.west);
  \end{tikzpicture}
	
  \caption{\textbf{Forecasting vulnerabilities:}
	probability of \CVE disclosure as a function of time and source code}
  \label{fig:holy_grail}
\end{figure}
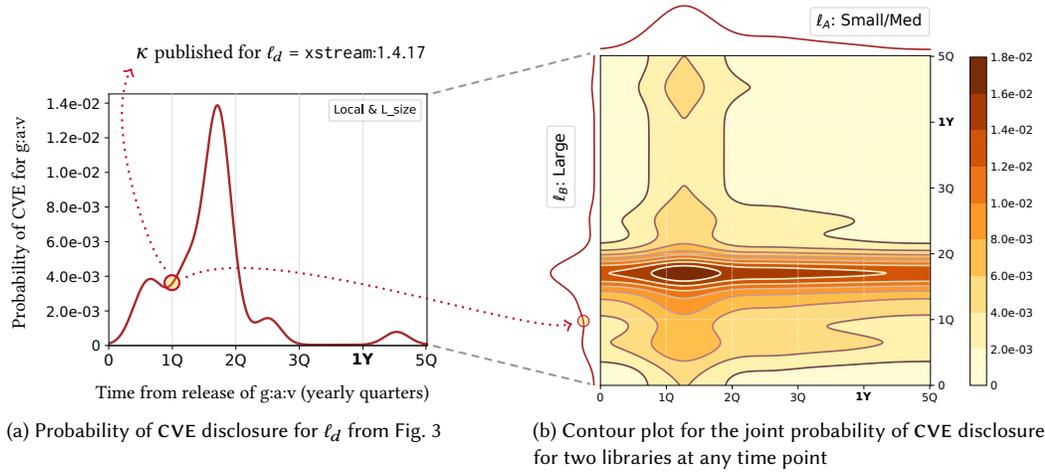

\Cref{fig:holy_grail} exemplifies the intention of \cref{obj:metric}: integrating the area under the \PDF in \Cref{fig:holy_grail:1Dplot} for $t_h$ days provides an estimate of the probability that the library instance faces a vulnerability in that time horizon.
As we show later in \Cref{sec:predict:method}, these metrics can be computed---for \FOSS projects---from data freely available online.


\smallskip
Going back to \Cref{ex:motivating_example}, hindsight shows that if \code{jira-core:8.19.1} had migrated to \code{xstream:1.4.18}, it would have not been vulnerable to \CVEn[1].
To know this \emph{before the disclosure of \CVEn[1]} one must quantify the probability of \CVE disclosure as a function of time and source code.
\Cref{fig:holy_grail:1Dplot} shows the quantification developed in this work applied to \code{xstream:1.4.17}.
By August, when version 8.19.1 of \lib[a] was being prepared for deployment, \code{xstream:1.4.17} was roughly three months old.
\Cref{fig:holy_grail:1Dplot} shows how the probability of \CVE disclosure for \lib[d] is then growing to its peak, meaning that our estimators advise a (fast) update to \code{xstream:1.4.18}---hindsight proved that this would have been the correct security move.

While such estimations are probabilistic, so they can ``prove right'' only a certain number of times, our intention is (not to guess the future but) to compute numeric indicators that help in making informed decisions.
That is shown in the previous example using \Cref{fig:holy_grail:1Dplot}---but it operates with a single dependency.
\Cref{fig:holy_grail:2Dplot} shows the same approach applied to two dependencies: one similar to \lib[d] from \Cref{ex:motivating_example}, and another one with a much smaller source code, say \lib[e] with $6k$ LoC.
This convolution of two probability density functions results in the probability plane depicted by the contour plot, where a \PDF is used as $y$-axis (this is the \PDF of \lib[d] in \Cref{fig:holy_grail:1Dplot}), and the other as the $x$-axis.

From such probability plane a \PDF can be extracted, equivalent to that in \Cref{fig:holy_grail:1Dplot}, but which describes the joint probability of a having a \CVE released for either \lib[d] or \lib[e], whichever happens first.
Thus, the generalization to $n$~dependencies proposed in \Cref{fig:holy_grail:2Dplot} is \emph{simple in theory, but hard to harness in practice} as it involves $n$-dimensional probability spaces whose computation requirements grow $\propto t_h^n$, i.e.\ exponentially on $n$.


\begin{hlbox}[RQ2:]
	How can one extend these forecasting metrics to software projects
	with dozens or hundreds of dependencies?
\end{hlbox}

This is addressed by our next objective:

\begin{objectives}[resume]
\item   \label{obj:scalable}
        Make the \PDF-based metrics scale polynomially on the number
		of codebases analyzed:
        \begin{itemize}[label=\textbullet,leftmargin=1.2em]
        \item   the aim is to process projects with dozens of dependencies
                within minutes in standard desktop systems;
        \item   moreover, while implementations can be language-specific,
                the concepts on which the metric builds should be
                applicable to any kind of open-source software.
        \end{itemize}
\end{objectives}

In \Cref{sec:model} we propose a succinct model for the evolution of software in time, that can be coupled with \PDF data via attack tree analysis, to produce the desired quantities in a time linear on the number of dependencies of the project.
In that sense, we posit that \emph{the dependency tree of a software project should be at the core of any prediction model, since the average commercial product includes \FOSS libraries that can make up 75\,\% of its codebase, and recent studies show that the total technical leverage can be much higher}~\cite{MP21b,Pit16}.
Therefore, a prime concern is to estimate which (\FOSS) dependencies will be responsible for most vulnerability disclosures---we discuss this further in \Cref{sec:background:related_work}, and in \Crefrange{sec:model:c-chains}{sec:model:TDT} we propose a way to study the problem at a high-level of abstraction, lightweight as per \cref{obj:scalable}.

However, while dependencies are key to reason about security vulnerabilities, standard dependency trees are snapshots that
do not capture the evolution of software in time.
Yet time plays a decisive factor in security, e.g.\ to identify the versions of a codebase that are affected by a vulnerability discovered in the latest release.

The literature usually takes a pragmatic approach to overcome such lack of expressiveness: when a vulnerability is disclosed for a library instance, it is common practice to consider vulnerable all prior versions as well---see \cite{MW10,BXHY22,NDM16} and \CVEn[1] in \Cref{fig:motivating_example}.
We argue that \emph{the main reason why traditional models for vulnerability analysis disregard software evolution is their focus on the vulnerabilities themselves, and not on the library instances where these are found, let alone their evolution in time}.
Our take is thus to enrich dependency trees with code versioning, merging code dependencies with software evolution, to allow for a simultaneous analysis of vulnerability propagation across \emph{the dependencies} and \emph{evolution in time} of a codebase.
The resulting concept of \emph{time dependency trees} introduced in \Cref{sec:model:TDT} allows to forecast the future vulnerabilities from any dependency of a library, spot codebases that act as single-point-of-failure for entire projects, quantify the health of an ecosystem in relationship to the number of libraries affected by vulnerability events, etc.

The above is one application example, but time dependency trees can serve further purposes.
They provide a minimal representation of the dependencies of a project across time, which allows for studying complete software environments to, among other things:
\begin{enumerate*}[itemjoin={{; }}]
\item	determine the presence of dependencies whose exploitation threatens seemingly independent projects
\item	spot codebases that represent a single-point-of-failure for several versions of a project
\item	quantify the health of an ecosystem in relationship to the number of library instances affected by vulnerability events, etc.
\end{enumerate*}

\subsubsection*{Article outline}
Next, after discussing related work and basic theory in \Cref{sec:background}, time dependency trees are introduced in \Cref{sec:model} and some general applications are outlined.
Then, \Cref{sec:predict} details the data process used to compute \PDFs like \Cref{fig:holy_grail:1Dplot}, and \Cref{sec:appMaven} shows its use on time dependency trees for the estimation of \CVE disclosure probabilities in Maven.
Finally, \Cref{sec:analyanal} describes a full-analytical alternative, and \Cref{sec:discussion} discusses the strong and weak points of all models and methods introduced.
This work concludes in \Cref{sec:conclu} and proposes some future lines of research.

\section{Background}
\label{sec:background}


\subsection{Related work on software vulnerabilities}
\label{sec:background:related_work}

There is an extensive body of work that studies software vulnerabilities, analyzing their discovery and prediction (i.e.\ foretelling its occurrence in the future) from measurable properties of software development projects.
To gather contributions relevant to this work, we systematised our literature search in Scopus and consulted the survey by \citet{GS17}; more details about the literature search are given in \Cref{sec:literature_research}.
We refined the search results by
focusing on works that introduce methods/models to discover or foretell vulnerabilities in code, including correlation studies between project metrics, and number and severity of vulnerabilities.
We classify and discuss the resulting 26 works---and the research gaps found---next and in \Cref{tab:related_work}.\!%

\begin{table}
	\centering
	\caption{Works that have proposed models to discover or foretell
	         vulnerabilities in source code}
	\label{tab:related_work}
	\vspace{-2ex}
	
	\def\colnote#1{\textsuperscript{\,#1\,}}
	\def\colnotetext#1#2#3{\hspace*{1em}\colnote{#1}{\textbf{#2}: #3}}
	\begin{smaller}
		\colortablepreamble
\def\CSEP{\hspace*{1.2em}}
\def\CWIDTH{.039\linewidth}
\def\CLWIDTH{.035\linewidth}
\def\HACKER{}
\def\yep{\checkmark}
\def\langs#1{\textsmaller{#1}}
\begin{tabular}{%
		         C{.028\linewidth}
		@{\CSEP} C{\CLWIDTH}  
		@{~~}    C{\CLWIDTH}  
		@{\CSEP} C{\CWIDTH}  
		@{~~}    C{\CWIDTH}  
		@{~~}    C{\CWIDTH}  
		@{~~}    C{\CWIDTH}  
		@{\CSEP} C{\CWIDTH}  
		@{~~}    C{\CWIDTH}  
		@{~~}    C{\CWIDTH}  
		@{\CSEP} C{\CLWIDTH} 
		@{~~}    C{\CLWIDTH} 
		@{~~}    C{\CLWIDTH} 
		@{\CSEP} L{.075\linewidth}  
		@{~}     R{0.04\linewidth}  
		@{\CSEP} L{.17\linewidth}
	}
	\toprule
	& \multicolumn{2}{c}{\textbf{\!\!\!Goal}\colnote{$\heartsuit$}}
	& \multicolumn{4}{c}{\textbf{Data}\colnote{$\clubsuit$}}
	& \multicolumn{3}{c}{\textbf{Method}\colnote{$\diamondsuit$}}
	& \multicolumn{3}{c}{\textbf{Approach}\colnote{$\spadesuit$}}
	& \multicolumn{2}{c}{\textbf{Projects/Libs.}\phantom{n}}
	& 
	  \multicolumn{1}{c}{\scalebox{.07}{\fontfamily{cmr}\selectfont\color{gray}%
		The rare-event nature of vulnerability disclosures
		usually passes unnoticed\HACKER}}
	\\
	\cmidrule[0pt]{1-1}
	\cmidrule[.3pt](r{5pt}){2-3}          
	\cmidrule[.3pt](l{1pt}r{5pt}){4-7}    
	\cmidrule[.3pt](l{1pt}r{5pt}){8-10}   
	\cmidrule[.3pt](l{1pt}r{5pt}){11-13}  
	\cmidrule[.3pt](l{1pt}r{3pt}){14-15}  
	\\[-3ex]
	\rowcolor{white}
		\multirow{-2.1}{*}{\rotatebox{90}{\bfseries Work}}
		& \rotatebox{45}{Disc.}
		& \rotatebox{45}{Pred.}
		& \rotatebox{45}{\CVEs}
		& \rotatebox{45}{Code}
		& \rotatebox{45}{\VCS}
		& \rotatebox{45}{Dep.}
		& \rotatebox{45}{Corr.}
		& \rotatebox{45}{Clas.}
		& \rotatebox{45}{T.Ser.}
		& \raisebox{.3ex}{AH}
		& \raisebox{.3ex}{SA}
		& \raisebox{.3ex}{ML}
		& \raisebox{.3ex}{Language}
		& \raisebox{.3ex}{\textlarger{\#}}
		& \multicolumn{1}{c}{\multirow{-1}{*}{\bfseries Purport}}
	\\
	\midrule
	\rowcolor{shade2}
		\cite{WTT+24}
		& \yep	
		&     	
		&     	
		& \yep	
		&     	
		&     	
		& \yep	
		& \yep 	
		&     	
		&     	
		&     	
		& \yep 	
		& \langs{C/\Cpp}
		& 20
		&
	\\
	\rowcolor{shade2}
		\cite{BES+20}
		& \yep	
		&     	
		&     	
		& \yep	
		&     	
		&     	
		&     	
		& \yep 	
		&     	
		&     	
		&     	
		& \yep 	
		& \langs{C}
		& 3
		&
	\\
	\rowcolor{shade2}
		\cite{AT17}
		& \yep	
		&     	
		&     	
		&     	
		& \yep	
		&     	
		& \yep	
		& \yep 	
		&     	
		&     	
		&     	
		& \yep 	
		& \langs{PHP}
		& 3
		&
	\\
	\rowcolor{shade2}
		\cite{BCH+14}
		& \yep	
		&     	
		&     	
		& \yep	
		& \yep	
		&     	
		&     	
		& \yep 	
		&     	
		& \yep	
		&     	
		&      	
		& \langs{C/\Cpp, PHP, Java, JS, SQL}
		& 10
		& \multirow{-4.1}{\linewidth}{%
			\raggedright\smaller%
			Find vulnerabilities regardless of existent logs such as \CVEs
			(although \acronym{cwe}{s} may be used).
			This includes formal methods and static/dynamic code analysis.}
	\\
	\rowcolor{shade1}
		\cite{LYZ+23}
		& \yep	
		&     	
		&     	
		& \yep	
		& \yep	
		&     	
		&     	
		& \yep 	
		&     	
		& \yep	
		&     	
		& \yep 	
		& \langs{C, Java}
		& 549
		&
	\\
	\rowcolor{shade1}
		\cite{LKKL14}
		& \yep	
		&     	
		& \yep	
		&     	
		& \yep	
		&     	
		&     	
		& \yep 	
		&     	
		& \yep	
		&     	
		&      	
		& \langs{C}
		& 3
		&
	\\
	\rowcolor{shade1}
		\cite{MSM+13}
		& \yep	
		&     	
		& \yep	
		&     	
		& \yep	
		&     	
		& \yep	
		&      	
		&     	
		& \yep	
		&     	
		&      	
		& \langs{C}
		& 1
		&
	\\
	\rowcolor{shade1}
		\cite{MW10}
		& \yep	
		&     	
		& \yep	
		&     	
		& \yep	
		&     	
		& \yep	
		&      	
		&     	
		& \yep	
		& \yep	
		&      	
		& \langs{C, ASM}
		& 3
		& \multirow{-4}{\linewidth}{%
			\RaggedRight\smaller%
			Detect known vulnerabilities (and their correlation to developer
			activity metrics) from \VCS only---e.g.\ commit churn,
			peer comments, etc.}
	\\
	\rowcolor{shade2}
		\cite{CKDR21}
		& \yep	
		&     	
		& \yep	
		& \yep 	
		&     	
		&     	
		&     	
		& \yep 	
		&     	
		&     	
		&     	
		& \yep 	
		& \langs{C/\Cpp}
		& 3
		&
	\\
	\rowcolor{shade2}
		\cite{GOP21}
		& \yep	
		&     	
		& \yep	
		& \yep 	
		&     	
		&     	
		&     	
		& \yep 	
		&     	
		&     	
		&     	
		& \yep 	
		& \langs{Java}
		& 7
		&
	\\
	\rowcolor{shade2}
		\cite{SAC21}
		& \yep	
		&     	
		& \yep	
		& \yep 	
		&     	
		&     	
		& \yep	
		& \yep 	
		&     	
		&     	
		& \yep	
		& \yep 	
		& \langs{Java}
		& 4
		&
	\\
	\rowcolor{shade2}
		\cite{SDW17}
		& \yep	
		&     	
		& \yep	
		& \yep 	
		&     	
		&     	
		& \yep	
		&      	
		&     	
		&     	
		& \yep	
		&      	
		& \langs{Java}
		& 3
		&
	\\
	\rowcolor{shade2}
		\cite{SW17}
		& \yep	
		&     	
		& \yep	
		& \yep 	
		&     	
		&     	
		& \yep	
		&      	
		&     	
		&     	
		& \yep	
		&      	
		& \langs{Java}
		& 5
		&
	\\
	\rowcolor{shade2}
		\cite{SMM+12}
		& \yep	
		&     	
		& \yep	
		& \yep 	
		&     	
		&     	
		&     	
		& \yep 	
		&     	
		& \yep	
		&     	
		&      	
		& \langs{C}
		& 7
		& \multirow{-6}{\linewidth}{%
			\RaggedRight\smaller%
			Detect known vulnerabilities (and their correlation to code
			metrics) from code only---e.g.\ number of classes, code cloning,
			cyclomatic complexity, etc.}
	\\
	\rowcolor{shade1}
		\cite{AL21}
		& \yep	
		&     	
		& \yep	
		& \yep	
		& \yep	
		&     	
		& \yep	
		& \yep 	
		&     	
		&     	
		&     	
		& \yep 	
		& \langs{C/\Cpp}
		& >150k
		&
	\\
	\rowcolor{shade1}
		\cite{KWLO17}
		& \yep	
		&     	
		& \yep	
		& \yep	
		& \yep	
		&     	
		&     	
		& \yep 	
		&     	
		& \yep	
		&     	
		&      	
		& \langs{C/\Cpp}
		& 8
		&
	\\
	\rowcolor{shade1}
		\cite{AFA16}
		& \yep	
		&     	
		& \yep	
		& \yep	
		& \yep	
		&     	
		& \yep	
		&      	
		&     	
		&     	
		& \yep	
		&      	
		& \langs{C/\Cpp}
		& 5
		&
	\\
	\rowcolor{shade1}
		\cite{CZ11}
		& \yep	
		&     	
		& \yep	
		& \yep	
		& \yep	
		&     	
		& \yep	
		& \yep 	
		&     	
		&     	
		& \yep	
		& \yep 	
		& \langs{C/\Cpp, Java}
		& 1
		&
	\\
	\rowcolor{shade1}
		\cite{SMWO11}
		& \yep	
		&     	
		& \yep	
		& \yep	
		& \yep	
		&     	
		& \yep	
		&      	
		&     	
		&     	
		& \yep	
		& \yep 	
		& \langs{C/\Cpp}
		& 2
		& \multirow{-5}{\linewidth}{%
			\RaggedRight\smaller%
			Detect known vulnerabilities (and their correlation to code
			and developer activity metrics) from both code and \VCS,
			but without considering the effect of dependencies in their
			propagation.}
	\\
	\rowcolor{shade2}
		\cite{PPP+22}
		& \yep	
		&     	
		& \yep	
		& \yep	
		& \yep	
		& \yep	
		&     	
		& \yep 	
		&     	
		& \yep	
		&     	
		&      	
		& \langs{Java}
		& 500
		&
	\\
	\rowcolor{shade2}
		\cite{LCF+22}
		& \yep	
		&     	
		& \yep	
		& \yep	
		& 	    
		& \yep	
		&     	
		& \yep 	
		&     	
		& \yep	
		&     	
		&      	
		& \langs{JS}
		& 624
		&
	\\
	\rowcolor{shade2}
		\cite{LST+21}
		& \yep	
		&     	
		& \yep	
		& \yep	
		&     	
		& \yep	
		&     	
		& \yep 	
		&     	
		&     	
		&     	
		& \yep 	
		& \langs{Java}
		& >300k
		&
	\\
	\rowcolor{shade2}
		\cite{PSS+21}
		& \yep	
		&     	
		& \yep	
		& \yep	
		& \yep	
		& \yep	
		& \yep	
		& \yep 	
		&     	
		&     	
		& \yep	
		&      	
		& \langs{Java, Ruby, Python}
		& 450
		& \multirow{-5}{\linewidth}{%
			\RaggedRight\smaller%
			Detect known vulnerabilities using code or \VCS, via
			de\-pend\-ency-\-aware models that can find the offending code,
			thus helping in its resolution (own vs.\ third-\-party library).}
	\\
	\rowcolor{shade1}
		\cite{LRW22}
		&     	
		& \yep	
		& \yep	
		&     	
		&     	
		&     	
		&     	
		&      	
		& \yep	
		&     	
		& \yep	
		& \yep 	
		& \langs{Agnostic}
		& 4
		&
	\\
	\rowcolor{shade1}
		\cite{YPWS20}
		&     	
		& \yep	
		& \yep	
		&     	
		&     	
		&     	
		&     	
		&      	
		& \yep	
		&     	
		& \yep	
		& \yep 	
		& \langs{Agnostic}
		& 9
		&
	\\
	\rowcolor{shade1}
		\cite{Las16}
		&     	
		& \yep	
		& \yep	
		&     	
		&     	
		&     	
		&     	
		&      	
		& \yep	
		&     	
		& \yep	
		& \yep 	
		& \langs{Agnostic}
		& 25
		&
	\\
	\rowcolor{shade1}
		\cite{RMR15}
		&     	
		& \yep	
		& \yep	
		&     	
		&     	
		&     	
		&     	
		&      	
		& \yep	
		&     	
		& \yep	
		&      	
		& \langs{Agnostic}
		& 5
		& \multirow{-4}{\linewidth}{%
			\RaggedRight\smaller%
		    Time regression to predict vulnerabilities from \acronym{nvd} logs,
		    but the models do not use domain-specific data relevant
			for security.}
	\\[.5ex]
	\bottomrule
\end{tabular}
\begin{tikzpicture}[overlay]
	\def\ATL{( -.876\linewidth,  .254\textheight )}
	\def\ATR{( -.848\linewidth,  .254\textheight )}
	\def\ABL{( -.876\linewidth, -.302\textheight )}
	\def\ABR{( -.848\linewidth, -.302\textheight )}
	\def\BTL{( -.676\linewidth,  .254\textheight )}
	\def\BTR{( -.648\linewidth,  .254\textheight )}
	\def\BBL{( -.676\linewidth, -.302\textheight )}
	\def\BBR{( -.648\linewidth, -.302\textheight )}
	\def\CTL{( -.520\linewidth,  .254\textheight )}
	\def\CTR{( -.492\linewidth,  .254\textheight )}
	\def\CBL{( -.520\linewidth, -.302\textheight )}
	\def\CBR{( -.492\linewidth, -.302\textheight )}
	\def\DTL{( -.780\linewidth, -.231\textheight )}
	\def\DTR{( -.646\linewidth, -.231\textheight )}
	\def\DBL{( -.780\linewidth, -.304\textheight )}
	\def\DBR{( -.646\linewidth, -.304\textheight )}
	%
	\draw [pattern={Lines[angle=-40,distance={4pt},line width=.8pt]},
	       pattern color=red,draw=none,rounded corners=3.5pt,opacity=.09,
	       yscale=.893,yshift=16.2pt]
		\ATL -- \ATR -- \ABR -- \ABL -- cycle;
	\draw [densely dashed,semithick,rounded corners=3.5pt,
	       draw=red!75!black!88,draw opacity=.8,
	       fill=red,fill opacity=.08]
		\ATL -- \ATR -- \ABR -- \ABL -- cycle;
	\node[rectangle,fit={\ATL \ABR},xshift=-2.7ex] {%
		\rotatebox{90}{\hspace*{-14.8em}\color{red!75!black!75}
		\sffamily\bfseries%
		Most works try to discover current vulnerabilities,
		not predict future ones}};
	%
	\draw [pattern={Hatch[distance={3pt},line width=.6pt,angle=405]},
	       pattern color=TealBlue,draw=none,rounded corners=3.5pt,opacity=.25,
	       yscale=.717,yshift=54pt]
		\BTL -- \BTR -- \BBR -- \BBL -- cycle;
	\draw [pattern={Hatch[distance={3pt},line width=.6pt,angle=40,yshift=3pt]},
	       pattern color=TealBlue,draw=none,rounded corners=2pt,opacity=.25,
	       yscale=.105,yshift=-1395pt]
		\BTL -- \BTR -- \BBR -- \BBL -- cycle;
	\draw [rounded corners=3.5pt,
	       draw=TealBlue,draw opacity=.8,
	       fill=TealBlue,fill opacity=.08]
		\BTL -- \BTR -- \BBR -- \BBL -- cycle;
	\node[rectangle,fit={\BTL \BBR},xshift=-2.7ex] {%
		\rotatebox{90}{\hspace*{-7.8em}\color{TealBlue!80!black}
		\sffamily\bfseries%
		Most works disregard the code dependency tree}};
	%
	\draw [pattern={Lines[angle=40,distance={4pt},line width=.8pt]},
	       pattern color=blue,draw=none,rounded corners=3.5pt,opacity=.09,
	       yscale=.893,yshift=16.2pt]
		\CTL -- \CTR -- \CBR -- \CBL -- cycle;
	\draw [dashdotted,semithick,rounded corners=3.5pt,
	       draw=blue!75!black!88,draw opacity=.8,
	       fill=blue,fill opacity=.08]
		\CTL -- \CTR -- \CBR -- \CBL -- cycle;
	\node[rectangle,fit={\CTL \CBR},xshift=-2.7ex] {%
		\rotatebox{90}{\hspace*{-14.8em}\color{blue!75!black!70}
		\sffamily\bfseries%
		Most works do not consider time in their analyses}};
	%
	\draw [pattern={Dots[angle=33,distance={4pt},radius=.7pt]},
	       pattern color=violet,draw=none,rounded corners=3.5pt,opacity=.12]
		\DTL -- \DTR -- \DBR -- \DBL -- cycle;
	\draw [densely dotted,very thick,rounded corners=3.5pt,
	       draw=violet,draw opacity=.5,
	       fill=violet,fill opacity=.1]
		\DTL -- \DTR -- \DBR -- \DBL -- cycle;
	\node[rectangle,fit={\DTL \DBR}] {%
		\,\parbox[t]{\linewidth}{\color{violet!70}%
		\sffamily\bfseries%
		Disregarded\\ security\\ data}};  
\end{tikzpicture}
%

	\end{smaller}

	\vspace{.5ex}%
	\smaller\RaggedRight
	\colnotetext{$\heartsuit$}{Goal}{%
		discovery/detection, or prediction (i.e.\ foretell
		occurrence in the future), of vulnerabilities in code.}\\
	\colnotetext{$\clubsuit$}{Data}{%
		main input of the model, namely source code, version control system,
		code dependency tree, or vulnerability records.}\\
	\colnotetext{$\diamondsuit$}{Method}{%
		Correlation (e.g.\ between churn and vulnerabilities),
		Classification (e.g.\ this function is vulnerable),
		or Time Series.}\\
	\colnotetext{$\spadesuit$}{Approach}{%
		Ad hoc (e.g.\ human inspection or custom scripts),
		Statistical Analysis, or Machine Learning.}\\
	\colnotetext{\;\;}{Note}{We omit works that do not propose new models
		to find/foretell vulnerabilities
		\cite{SCX+24,PBX+23,SXL+23,WLX+23,YLF+23,BXHY22,ITW21,KEA21,MP21a,DRW+18,GS17}.}
\end{table}
%
%

\newcommand{\tablecolumn}[2][]{\textsbf{#2}\ifthenelse{\equal{#1}{}}{}{\raisebox{1pt}{\smaller[2]$\mkern-3mu{/}\mkern3mu$}{#1}}\xspace}

\textbf{Machine Learning} (\ML) is today the most popular approach to studying software vulnerabilities, in terms of the number of projects and libraries analyzed.
This is followed by classical statistical analyses
such as hypothesis tests which, alongside \ML, count with well-known software support, for example, TensorFlow (Python, Java, \Cpp), SciPy (Python), Matlab models, the \textsf{R} programming language, etc.
Almost all these works present methods to analyze the source code or version control system (\VCS) of a library, to detect vulnerable code fragments or compute metrics that suggest its presence today.
In other words, they try to \emph{discover existent vulnerabilities}---by applying available algorithms to massive amounts of data---but disregard the \emph{estimation of future vulnerabilities}, as indicated by the blanks in the columns \tablecolumn[Pred.]{Goal} and \tablecolumn[T.Ser.]{Method} in \Cref{tab:related_work}.
Therefore, even though the proposed model %
have a varying degree of accuracy, and leaving aside the known limitations on the explainability of the outcomes \cite{GS17}, these models do not study the future consequences of current choices (e.g.\ a decision to keep legacy dependencies), which are known to affect the number, severity, and type of vulnerabilities faced in the future \cite{YPWS20,PPP+22}.

\begin{gap}[Foretelling vulnerabilities]
	\label{gap:foretelling_vulnerabilities}
	Literature in the field focuses on existent vulnerabilities---%
	estimating the (likelihood of) occurrence of vulnerabilities
	in the future is a largely neglected field.
\end{gap}

A smaller set of works train time regression models on \NVD data, addressing \Cref{gap:foretelling_vulnerabilities} to some extent \cite{LRW22,YPWS20,Las16,RMR15}.
Akin to the methods used for stock market fluctuations---e.g.\ \acronym{arima}---these works forecast the number of vulnerabilities expected in software projects.
This can help managerial decisions to a certain degree, e.g.\ to draw a ballpark figure of the global manpower needed for project security updates.
However, such high-level predictions cannot give deeper insight into the nature of the vulnerabilities \cite{MN14,DBM18}.
This is unavoidable in time regression models trained exclusively on the variable to predict---number of \CVEs in this case---that ignore other data required for domain-specific forecasting, as highlighted with a dotted box in the last three rows of \Cref{tab:related_work}.

\begin{gap}[Coarse prediction models]
	\label{gap:coarse_prediction_models}
	The relatively few works that attempt to foretell future vulnerabilities
	do so at a high level---they give no white-box, low-level information
	that can help in its mitigation.
\end{gap}

A vulnerability forecast, or discovery for that matter, is most useful when it provides information on the nature and causes of the problem, that developers can use to fix them or mitigate its impact.
That is why most works in \Cref{tab:related_work} are project- or language-specific, using code metrics or \VCS data to pinpoint which functions or commits are likely to contain a vulnerability.
In that sense, the dependency tree of a library provides vital information regarding inherited vulnerabilities, which could make up to 75\,\% of the total \cite{Pit16}.
However, and as highlighted in the column \mbox{\tablecolumn[Dep.]{Data}} of \Cref{tab:related_work}, only four works were found to use the dependency tree to discover vulnerabilities in libraries, and we found no work that uses it to foretell vulnerabilities \cite{LCF+22,LST+21,PSS+21,PPP+22}.

\begin{gap}[Code dependency tree]
	\label{gap:code_dependency_tree}
	The dependency tree is key to analyse the vulnerabilities of a library---%
	however, all vulnerability-foretelling works disregard it,
	and very few vulnerability-discovery works make use of it.
\end{gap}

The dependency tree is paramount to foretell vulnerabilities, since the typical commercial product includes libraries for up to three-quarters of its codebase, thus constituting a major source of potential attacks \cite{Pit16,KGO+18,PPP+22}.
From such future vulnerabilities, the biggest threat is posed by those which will be \emph{publicly disclosed}, as these enable immediate attacks and trigger urgent and costly remediation steps.
%

Most works in \Cref{tab:related_work} look for such threats, leveraging \CVE logs to detect or predict vulnerable code fragments.
In other words, their granularity is finer than files, which has resulted in limited project- and time-extrapolation~\cite{PSS+21}.
Instead, we focus on the coarser level of \emph{vulnerable library instances}, by using \NVD data to identify instances in the dependency tree that are affected by disclosed vulnerabilities, which are the red flags whose exclusion pays off to engineer \cite{KGO+18,HDG18}.
Besides addressing the generalisability of the forecasts, this is coherent with the fact that most of the codebase is not controlled by the project team.
Therefore, it is arguably more realistic to engineer the change of a dependency (e.g.\ update/drop a library instance), than to expect to be able to modify its source code \cite{MP21b}.

But such coarsening in the level of analysis does not change the fact that vulnerabilities (or vulnerable library instances) are a minority of code when compared to the complete codebase of a project.
This brings forward another research shortcoming: the rarity of vulnerability disclosures, which is overlooked by most of the literature.
%
%
But disregarding rare events is not possible in a field where a missing bounds check affecting 40 LoC endangers millions of credit accounts%
\footnote{CVE-2014-0160 ``Heartbleed'' \cite{CDFW14}: \href{https://github.com/openssl/openssl/commit/731f431497f463f3a2a97236fe0187b11c44aead?diff=split}{\ttfamily https://github.com/openssl/openssl/commit/731f431497f463f3a2a97236fe0187b11c44aead?diff=split}}\!.
What happens instead is that most models that try to discover and foretell vulnerabilities, treat these as usual (not rare) events.
The issue with this is that, from a modeling perspective, it is fundamental to account for such rare nature of the input data, since generalizing from niche cases---e.g.\ two vulnerable code fragments in a codebase with hundreds of files, each with dozens of versions---is bound to bias most estimation models \cite{YPWS20}.

\begin{gap}[Vulnerabilities are rare events]
	\label{gap:vulnerabilities_are_rare_events}
	The expected number of vulnerabilities affecting a library in an update
	period is zero---however, most works disregard this, deploying models
	which therefore have limited extrapolation capacity.
\end{gap}

Only two works in \Cref{tab:related_work} acknowledge \Cref{gap:vulnerabilities_are_rare_events}:
\citet{SMWO11} mention that \textquote{vulnerabilities are rare occurrences} but their methodology shows no dedicated rare-event treatment;
\citet{YPWS20} interpreted vulnerability data as a time series---similarly to what we do in \Cref{sec:model:c-chains}---and used statistical regression models that would not suffer too much in predictive capability due to the rareness of vulnerabilities, e.g.\ Croston's.
This treats data uniformly, as in time series analysis for the stock market.
However, for software security, \citet{CKDR21} note that such models can be a hindrance since they \textquote{do not take into account semantic dependencies that play a vital role in vulnerability predictions}.

In contrast to the above, our approach to \Cref{gap:vulnerabilities_are_rare_events}---which for us translates into a scarcity of fitting data---is clustering.
Prior to statistical fitting of probability density functions (\PDFs), we lump together libraries whose properties result in a similar likelihood of vulnerability disclosures.
These ``properties'' are metrics from the code that are relevant for security, such as the base of external dependencies of a library as measured by its technical leverage \cite{MP21b}.
Thus, we address the rarity of vulnerabilities and at the same time mitigate overfitting, by proposing a model for the estimation of future vulnerabilities that is sensitive to (publicly available) semantic information of software security.

Regarding 
\crefrange{gap:foretelling_vulnerabilities}{gap:code_dependency_tree},
our approach in \Cref{sec:model,sec:predict} covers
the first two
by proposing a model to estimate future vulnerabilities, that can pinpoint the library instances where these may originate.
This makes use of the code dependency tree of the main library under analysis, thus covering \Cref{gap:code_dependency_tree}.

The end result are graphs designed to foretell vulnerabilities, which are white-box models of project code that developers can use to select dependencies, in a manner that minimises the vulnerabilities to expect in a foreseeable time horizon.
The theoretical foundations for the structure of these graphs, and for the propagation of vulnerability probabilities, is based on attack tree analysis.

\subsection{Attack tree analysis}
\label{sec:background:AT}

An attack tree (\AT) is a hierarchical diagram that describes potential attacks on a system.
Its single root on top represents the attacker's goal, and the leaves---known as \emph{basic attack steps} or \acronym{bas}---represent the indivisible actions of the attacker.
Intermediate nodes are labeled with gates that determine how their lower-connected nodes activate them: standard \ATs have \OR and \AND gates only, but many extensions exist to model more elaborate attacks \cite{KPC14}.

\begin{figure}
  \begin{minipage}{.5\linewidth}
	\centering
	\includegraphics[width=.98\linewidth]{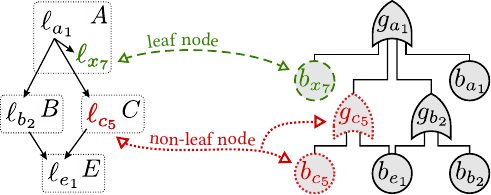}
	\captionof{figure}{Bijection between dependency trees (left) and \ATs}
	\label{fig:background:AT}
  \end{minipage}
  \hfill
  \begin{minipage}{.45\linewidth}
	\begin{example}
	\label{ex:dependency_tree_to_attack_tree}
		\Cref{fig:background:AT} to the left demonstrates the bijection
		between dependency trees and standard attack trees.
		There are two cases to distinguish:
		\begin{itemize}[leftmargin=1em]
		\item	{\LibertinusSerifSB leaf nodes}
				in the dependency tree are mapped to a single \BAS
				in the \AT---see e.g.\ the dashed line in
				\Cref{fig:background:AT} for the leaf node
				$\lib[x_7]\mapsto \{b_{x_7}\}$;
		\item	{\LibertinusSerifSB non-leaf nodes}
				in the dependency tree are mapped to an \acronym{or} gate,
				and also to a \BAS---see e.g.\ the dotted line
				for the non-leaf node
				$\lib[c_5]\mapsto \{g_{c_5}, b_{c_5}\}$.
		\end{itemize}
 	\end{example}
  	\vspace{-2ex}
  \end{minipage}
  \vspace{-2ex}
\end{figure}

\ATs are closely related to our modeling of library attacks by exploits in the dependencies.
Structurally, they are equivalent to the dependency trees defined in \Cref{sec:rq:terminology}: the \AT root represents an attack on the main library, and the \acronym{bas}{es} represents compromising a dependency.
Nodes in \ATs have a type:
since a single vulnerable dependency among many can compromise a library, here it suffices to consider \OR intermediate nodes alone.
This results in the following bijection between a dependency tree ($D$) and an \AT (\thisAT), also illustrated in \Cref{fig:background:AT,ex:dependency_tree_to_attack_tree}:
\begin{itemize}[topsep=.5ex]
\item	each leaf \lib in $D$ maps to a \acronym{bas} $b_{\lib}$ in \thisAT,
		representing an exploit to the own code of \lib;
\item	each non-leaf node $\lib'$ in $D$ maps to an \OR gate $g_{\lib'}$
		and a \acronym{bas} $b_{\lib'}$ in \thisAT such that:
		\begin{itemize}
		\item	$b_{\lib'}$ represents an exploit to the own code of $\lib'$;
		\item	$g_{\lib'}$ has an input edge from $b_{\lib'}$;
		\item	each direct dependency of $\lib'$ in $D$ is an input edge
				of $g_{\lib'}$.
		\end{itemize}
\end{itemize}

Besides their similar structure, \ATs are interesting for this study because they are often analyzed quantitatively to compute security metrics.
Typical examples are the minimal time \cite{AHPS14}, minimal cost \cite{AGKS15}, or maximal probability \cite{JKM+15} of a successful attack, as well as Pareto analyses that study trade-offs among attributes \cite{FW19}.
\emph{We are mostly interested in probability propagation, in the \AT representation of a dependency tree, which we will use to compute the time-bounded likelihood of a vulnerability exploit.}
Further metrics can be approached similarly, e.g.\ estimating the expected time to next zero-day by representing the occurrence of security vulnerabilities with rates of \CVE publication.

\subsection{Common assumptions in the field}
\label{sec:background:assumptions}

In what follows we work under the standard assumptions of the literature in the field---see e.g.\ the works in \Cref{tab:related_work}:
\begin{assumptions}
\item	\label{ass:similarlib->similarvuln}
		In a given time span, libraries with similar (security-relevant)
		properties are likely to face a similar amount of vulnerability
		disclosures that affect them.
\item	\label{ass:pastvuln->futurevuln}
		The history of vulnerabilities faced by a library---or family
		of libraries---is indicative of the expected vulnerabilities
		to face in the future.
\end{assumptions}

\Cref{ass:similarlib->similarvuln} depends on \emph{time} and \emph{library properties}, where the former is typically approached under the principle of regression to the mean, viz.\ the longer the time span, the more likely it is to observe coincidences in the number of vulnerabilities observed for two libraries with similar ``properties''.

The concept of \emph{library properties} that are relevant for security purposes is less well understood, as elaborated in detail above.
\Cref{sec:predict:method} motivates the use of two such properties for our demonstration: \emph{own-size} and \emph{class} of the library, the latter divided between libraries with a web-oriented purpose (\code{Remote\:network}) or not (\code{Local}).
Different properties could be used---in any case, and in contrast to most related works in \Cref{tab:related_work}, our model does not use them directly to predict vulnerabilities.
Instead, we use properties to lump together libraries that can be considered similar from a security perspective, thereby creating clusters from which we then develop our prediction models, namely \PDFs.

This procedure aligns well with \cref{ass:pastvuln->futurevuln}, on which we stand to develop statistical fittings similar to works \cite{YPWS20,Las16}.
Two main differences with respect to such works are that (a) besides vulnerability records we also use domain knowledge, namely code metrics, and (b) we fit an independent model to each cluster of libraries, so different best-fit models can intervene in the estimation of vulnerabilities for one main library.

Thus, for us, \emph{clustering}---i.e.\ lumping together libraries with similar security properties---serves two purposes:
\begin{itemize}[topsep=.3ex,parsep=0pt,leftmargin=1.1em]
\item	\emph{mitigate rare events}, because zero vulnerabilities are expected
		to affect the typical library instance in a few months' time, so
		constructing an empirical model from the history of a cluster of
		libraries (as opposed to a single library) increases the amount
		of available data, and thus the quality of the model;
\item	\emph{avoid overfitting}, since the statistical or \ML algorithms
		are trained on a variety of data, expected to be similar
		from the security perspective but different in other dimensions.
\end{itemize}


\section{Evolution of library dependencies in time: a formal model}
\label{sec:model}

We seek a general and extensible way to estimate the probability of future vulnerabilities in our code, that may come from library dependencies.
This \namecref{sec:model} elaborates a formal basis for the general model we develop to that end---in \Cref{sec:predict} we show how to implement one such prediction model in practice.

\subsection{Evolution of a software codebase}
\label{sec:model:c-chains}

We work with a continuous notion of time.
For (time) interval $T\subsetneq\RR$ we use square brackets and parentheses respectively to signify closed and open ends, e.g.\ $T = [\tstart,\tend)\neq\emptyset$ means that $\tstart\in T$ and $\tend\notin T$.
When a new notation is introduced we highlight it by using the symbol \,$\doteq$\, instead of equals (${=}$).

It is standard to assume a total order of versions in the development of a library in time---see \citet{SMWO11,PPP+18,PSS+21,PPP+22}, etc.
This disregards situations when a chain of releases is forked, and development is maintained for two or more henceforth incomparable versions of a library.

\begin{example}
\label{ex:apache_tomcat}
	Apache Tomcat offers parallel support for versions 8.5, 9.0, and 10.0
	\cite{Tomcat}, which results in there not being a total order among
	such versions of the library.
	For instance, code changes uploaded to version
	\href{https://github.com/apache/tomcat/commit/b5c9c3a25a17f777989408973013f5312acdb8e2}{8.5.75} (Jan 17, 2022) were not added to version \href{https://github.com/apache/tomcat/commit/bd9afafc1ec568f8160ed3679a776b26d8a29b99}{9.0.58} (Jan 15, 2022).
\end{example}

To cover these cases, we consider each parallel version of the library as a ``codebase chain'' independent from the rest.
In \Cref{ex:apache_tomcat} this means that Tomcat\;v8.5 defines a chain of development, and Tomcat\;v9.0 defines a different one.
This solution allows modeling common patches as parallel updates on each chain, which is a common practice e.g.\ via git cherry-picking.
Moreover, notwithstanding the commit history, the released version-tagged codebases (i.e.\ library instances%
\footnote{%
This assumes a bijection between CVS tags (e.g.\ in GitHub) and repository library versions (Maven), but some scenarios can be more complex \cite{HSCT21}.
}%
) are comparable among each other \emph{within a chain}.
We formalise all this under the concept of c-chains.

\begin{definition}[c-chain]
\label{def:c-chain}
	A \emph{codebase chain}, or \emph{c-chain} for short,
	is a series $\left\{\lib[v_j]\right\}_{j=1}^m$ of library instances
	such that there is a total order ${<}$ for $\{v_j\}_{j=1}^m$
	based on their release dates, where $v_{j'}>v_j$ indicates that
	\lib[v_{j'}] was developed from the codebase of \lib[v_j].
\end{definition}

We use the abbreviated notation $\lib[v_1] < \lib[v_2] < \cdots < \lib[v_m]$ to denote the c-chain of library \lib with $m$ library instances, $\left\{\lib[v_j]\right\}_{j=1}^m$, whose versions satisfy $v_j<v_{j'}$ for all $1\leqslant j < j' \leqslant m$.

\Cref{def:c-chain} allows for a single library \lib to span multiple (incomparable) c-chains.
In such cases, we identify each independent c-chain by appending some unique id---e.g.\ its \code{MAJOR.Minor} version number---to the library name/symbol.
This singles out the c-chain, which can then have the standard version subindices described in \Cref{sec:rq:terminology}.

\begin{example}
\label{ex:apache_tomcat_c-chain}
	Applying this concept to \Cref{ex:apache_tomcat} on Apache Tomcat yields
	e.g.\
	\begin{itemize*}[topsep=0pt,label=,itemjoin={{, }},itemjoin*={{, and}}]
	\item	
		$\clib[8.5.74]{8.5} < \clib[8.5.75]{8.5} < \clib[8.5.76]{8.5}$
		for the c-chain of v8.5
	\item	
		$\clib[9.0.57]{9.0} < \clib[9.0.58]{9.0} < \clib[9.0.59]{9.0}$
		for v9.0.
	\end{itemize*}
	Since elements are incomparable across c-chains, neither $\clib[8.5.75]{8.5} < \clib[9.0.58]{9.0}$ nor $\clib[8.5.75]{8.5} > \clib[9.0.58]{9.0}$ (denoted $\clib[8.5.75]{8.5} \nlgt \clib[9.0.58]{9.0}$), in accordance to \Cref{ex:apache_tomcat}.
\end{example}

This resembles modeling \clib{8.5} and \clib{9.0} as independent libraries.
The main difference is that c-chains of a library almost always belong to the same project, which affects metrics like technical leverage \cite{MP21b}.
Furthermore, library instances from different c-chains of the same library are likely to exhibit correlated---or even the same---vulnerabilities.
That is, even though their code evolution is incomparable, the common origin of the main codebase results in a relatively high probability of common vulnerabilities.
For concrete examples consider \href{https://www.cve.org/CVERecord?id=CVE-2021-33037}{CVE-2021-33037} and \href{https://www.cve.org/CVERecord?id=CVE-2021-41079}{CVE-2021-41079}, both affecting various library instances in the c-chains v8.5, v9.0, and v10.0 of Apache Tomcat.


This parallel development, but the likelihood of common vulnerabilities, is key to estimating the disclosure of vulnerabilities in the future.
Below in \Cref{sec:model:TDT} we introduce a model to facilitate such estimation, by considering the dependency tree of a library, articulated with the evolution of software in time via structures derived from c-chains.
Ideally, this would simply use c-chains (of dependencies) to link in time the various dependency trees of some main library.
However, that cannot model different instances of a library coexisting in a dependency tree---we revise that next.

\subsection{Different instances of the same library in a dependency tree}
\label{sec:model:d-matrices}

While c-chains can capture the evolution of a library in isolation, they fall short of modeling its embedding in a software ecosystem.
The main modeling limitation to consider is when a dependency tree has different instances of one library.

\begin{figure}
	\centering
	\begin{subfigure}[b]{.4\linewidth}
		\centering
		\includegraphics[width=.8\linewidth]{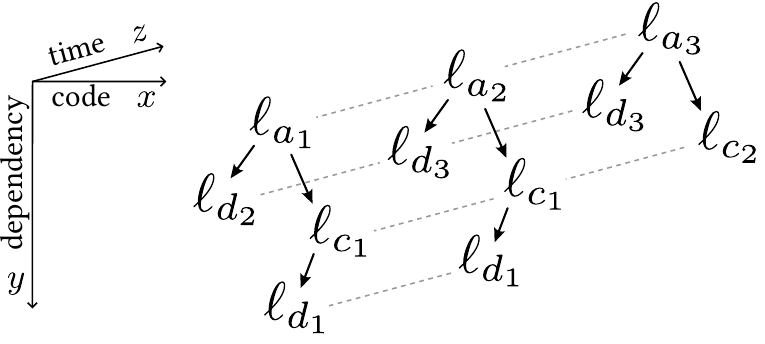}
		\caption{Dependency tree of \lib[a] in three consecutive versions}
		\label{fig:d-matrices:dependency_trees}
	\end{subfigure}
	\quad
	\begin{subfigure}[b]{.25\linewidth}
		\centering
		\begin{align*}
			\lib[d_2] < \lib[d_3]  &= \bflib[d]{[1]} 
			\\
			\lib[d_1]              &= \bflib[d]{[2]} 
		\end{align*}
		\caption{c-chains of library \lib[d]}
		\label{fig:d-matrices:c-chains}
	\end{subfigure}
	\quad
	\begin{subfigure}[b]{.25\linewidth}
		\centering
		\[
			\begin{bmatrix}
				\lib[d_2] & \lib[d_3] & \lib[d_3]\\
				\lib[d_1] & \lib[d_1] & \libdummy
			\end{bmatrix}
			\doteq
			\bflib[d]
		\]
		\caption{d-matrix of library \lib[d]}
		\label{fig:d-matrices:d-matrix}
	\end{subfigure}
	\caption{Tracking different instances of the library \lib[d] in the
	         dependency tree of the library \lib[a] across multiple versions}
	\label{fig:d-matrices}
\end{figure}

\begin{example}
\label{ex:multiple_instances_of_a_dependency}
	Library \lib[d] in \Cref{fig:d-matrices:dependency_trees}
	is a dependency of the root library \lib[a].
	The leftmost dependency tree in \Cref{fig:d-matrices:dependency_trees}
	shows that \lib[d_2] is a direct dependency of \lib[a_1],
	and at the same time \lib[d_1] is a transitive dependency (via \lib[c_1]).
	These are two different versions of \lib[d], which means that
	the root \lib[a_1] simultaneously depends on both instances of
	that library: \lib[d_2] and \lib[d_1].
	Later on, the developers of \lib[a] release a new version---\lib[a_2],
	see the dependency tree in the middle of
	\Cref{fig:d-matrices:dependency_trees}---updating its direct dependency
	to \lib[d_3]. Therefore, at that next point in time, the root library
	depends on \lib[d_3] and \lib[d_1].
\end{example}

In general, when studying the evolution in time of a dependency tree, a dependency like \lib[d] in \Cref{ex:multiple_instances_of_a_dependency} gives rise to not just one, but many c-chains.
In the \namecref{ex:multiple_instances_of_a_dependency} we have the c-chain $\lib[d_2]<\lib[d_3]$ on the one side (for \lib[d] as a direct dependency of \lib[a]), and \lib[d_1] alone on the other side (as a transitive dependency via \lib[c]).

To cater for these cases, one must keep track of all versions of each dependency that are present in a dependency tree.
Moreover, the adoption of new dependencies must be considered, as well as the dropping of deprecated ones.
We capture all these scenarios with the concept of d-matrix.

\begin{definition}[d-matrix]
\label{def:d-matrix}
	A \emph{dependency matrix}, or \emph{d-matrix for short},
	is a series $\bflib = \left\{\bflib[i]\right\}_{i=1}^n$ of $m$-tuples
	such that its $i$-th projection for $1\leqslant i \leqslant n$
	is a c-chain of library \lib,
	denoted $\bflib{[i]} \doteq \bflib[i] = \left\{\lib[j]\right\}_{j=1}^m$.
\end{definition}

D-matrices model the evolution in time of multiple dependencies of one library, as they may occur in a dependency tree.
For notational convenience we allow \emph{version stuttering}, e.g.\ the projection $\bflib{[i]}$ could yield the sequence of library instances \tuple{\lib[v_1], \lib[v_2], \lib[v_2], \lib[v_3]}, which is in fact the c-chain $\lib[v_1]<\lib[v_2]<\lib[v_3]$.
Version downgrading is treated analogously.

To model the adoption and dropping of dependencies we introduce the dummy symbol~\libdummy\!, used to represent the absence of a library instance in a d-matrix.
Symbol \libdummy would typically occur in the extremes of the projections of d-matrices, as in \tuple{\libdummy, \libdummy, \lib[v_1], \lib[v_2], \lib[v_2], \lib[v_3], \libdummy}:
here it represents a case in which, for the first two time steps, the library \lib was not a dependency of the tree under analysis;
then the library instance \lib[v_1] becomes a dependency;
then that dependency is updated to \lib[v_2], and kept like that for the next version;
then it is updated to \lib[v_3];
and finally the dependency is dropped for the last time step.
This too results in the c-chain $\lib[v_1]<\lib[v_2]<\lib[v_3]$.

\begin{example}
\label{ex:d-matrix}
	\Cref{fig:d-matrices:dependency_trees} shows how the root library \lib[a]
	depends on two different versions of \lib[d].
	The corresponding $2\times3$ d-matrix for that scenario is shown in
	\Cref{fig:d-matrices:d-matrix}, which exhibits version stuttering for
	instances \lib[d_3] and \lib[d_1], and uses the dummy library \libdummy
	to represent how \lib[c_2] dropped its dependency of \lib[d] in the
	last time step.
	The first and second projections of that d-matrix,
	\bflib[d]{[1]} and \bflib[d]{[2]}, are the c-chains of \lib[d] shown in
	\Cref{fig:d-matrices:c-chains}.
\end{example}

To keep a consistent order of library instances, the c-chains that form a d-matrix are aligned in time.
In other words, a d-matrix \bflib is structured like a matrix as in \Cref{fig:d-matrices:d-matrix}, with a row $1\leqslant i\leqslant n$ for each c-chain, and a column $1\leqslant j\leqslant m$ for each version of the main library under study.
Projecting the d-matrix by row alone yields a c-chain: $\bflib{[i]} = \bflib{[i,\cdot]} = \lib[1]<\lib[2]<\cdots<\lib[m]$.
Projecting by both row and column yields a library instance: $\bflib{[i,j]}=\lib[j]$.
As in \Cref{ex:multiple_instances_of_a_dependency}, some c-chains may be longer than others: this is catered for by taking the earliest and latest instances of the library as the extremes of all c-chains, and padding with \libdummy where necessary.


\smallskip
We now have the tools to define a formal model for the evolution of dependency trees in time: \emph{time dependency trees}.

\subsection{Time dependency trees}
\label{sec:model:TDT}

Standard dependency trees are time agnostic, since their nodes are library instances connected solely by code-dependency relationships.
This hinders the estimation of future vulnerabilities:
first, because such trees do not connect different instances from the same library (which may face positively-correlated vulnerability disclosures);
and, more importantly, because the development history of a library---ignored by standard dependency trees---has a bearing on the likelihood of incoming vulnerability discoveries.
Therefore, to develop a vulnerability-forecasting model, we see the need to merge standard dependency trees with a formalised notion of software evolution in time.

\begin{hlbox}[Time dependency trees]
	use d-matrices to give a structured time dimension to dependency trees,
	allowing for a simultaneous analysis of vulnerability propagation across
	\emph{the dependencies} and \emph{evolution in time} of a codebase.\!%
	\footnotemark
\end{hlbox}
\footnotetext{%
Possible connections to abstract interpretation theory and graph transformations are discussed below in \Cref{sec:conclu:future}.
}

\begin{definition}[\TDT]
\label{def:TDT}
	Let $\left\{\lib[v_j]\right\}_{j=1}^m\neq\emptyset$ be a c-chain;
	we call \emph{dependency tree of \lib[v_j]}, denoted \DepTree{\lib[v_j]},
	to the dependency tree whose root is the library instance \lib[v_j].
	Let also $\DepTree[t]{\lib} \doteq \DepTree{\lib[v_j]}$
	for any time $t \in \big[ \trel(\lib[v_j]), \trel(\lib[v_{j+1}]) \big)$,
	i.e.\ any time instant at or after the release of \lib[v_j],
	but before the release of \lib[v_{j+1}].
	Moreover let $T=\left[\tstart,\tend\right]$ be a time span that encloses
	the c-chain $\left\{\lib[{v_j}]\right\}_{j=1}^m$,
	such that the release times satisfy
	$\trel(\lib[{v_1}])=\tstart\leqslant\tend=\trel(\lib[{v_m}])$.
	\\
	The \emph{time dependency tree (\TDT) of library \lib
	in the time period $T$} is the family of dependency trees
	$\TimeDepTree[T]{\lib} \doteq \left\{\DepTree[t]{\lib}\right\}_{t\in T}$.
\end{definition}

When $T$ is clear from context we omit it: then the \TDT denoted $\TimeDepTree{\lib}\doteq\TimeDepTree[T]{\lib}$ is an infinite family which, instantiated at time $t\in T$, results in the standard dependency tree \DepTree[t]{\lib}.
This infinite semantics stems from our use of continuous time; nevertheless, the syntactic representation of a \TDT is finite and, moreover, minimal for the set of codebases considered.
More specifically, a \TDT can be represented as a \DAG such that:
\begin{enumerate}
\item	\label{TDT:node}
		there is one node per library instance,
		in the family of dependency trees within the time span $T$;
\item	\label{TDT:edge:dependency}
		there is an edge from each library instance to each direct dependency;
\item	\label{TDT:edge:c-chain}
		there is an edge from each library instance to any (one) successor
		in the corresponding c-chain.
\end{enumerate}

\begin{figure}
	\centering
	\begin{subfigure}[t]{.25\linewidth}
		\centering
		\includegraphics[width=\linewidth]{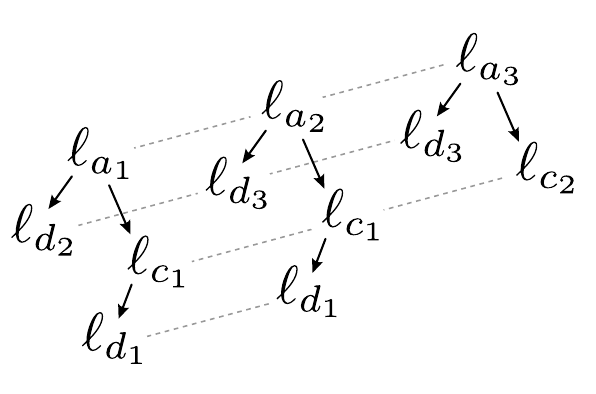}
		\caption{\parbox[t]{.8\linewidth}{%
		Dependency trees in time (same as \Cref{fig:d-matrices:dependency_trees})}}
		\label{fig:TDTs:TDT_example_0}
	\end{subfigure}
	\hfill
	\begin{subfigure}[t]{.25\linewidth}
		\centering
		\includegraphics[width=\linewidth]{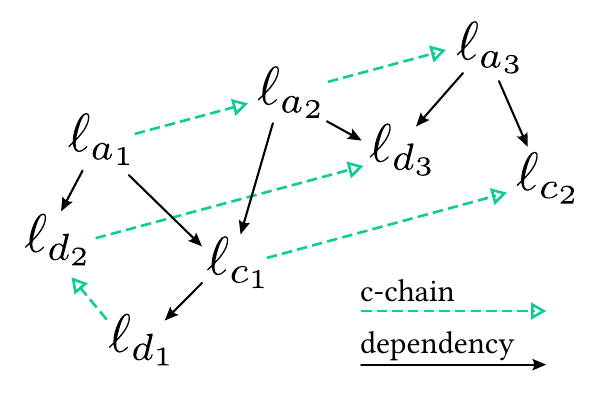}
		\caption{\TimeDepTree{\lib[a]} = \TDT of
		         \Cref{fig:TDTs:TDT_example_0} (left)}
		\label{fig:TDTs:TDT_example_1}
	\end{subfigure}
	\hfill
	\begin{subfigure}[t]{.35\linewidth}
		\centering
		\includegraphics[width=\linewidth]{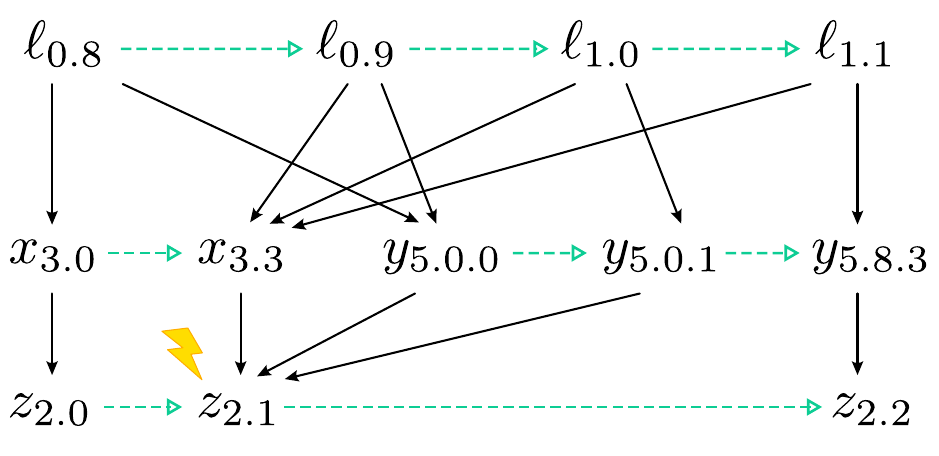}
		\caption{\TDT of four versions of \lib with \SPOF $z_{2.1}$}
		\label{fig:TDTs:TDT_example_2}
	\end{subfigure}
	
	\vspace{-2ex}
	\caption{Time Dependency Trees as per \Cref{def:TDT},
	         illustrative examples}
	\label{fig:TDTs}
\end{figure}

\begin{example}
\label{ex:TDT:time-indexing}
	Consider the sequence of dependency trees from \Cref{fig:TDTs:TDT_example_0},
	where the main library is \lib[a] and the implicit time span is
	$T=[t_1,t_3]$ for $t_i\doteq\trel(\lib[a_i])$.
	The corresponding \TDT of \lib[a], \TimeDepTree{\lib[a]},
	can be represented as the \DAG in \Cref{fig:TDTs:TDT_example_1},
	where dependency edges---item~(\ref{TDT:edge:dependency}) in the
	enumeration above---are black solid arrows, and c-chain edges---%
	item~(\ref{TDT:edge:c-chain})---are green dashed arrows.
	Instantiating \TimeDepTree{\lib[a]} at any time point before the release
	of \lib[a_2], $t\in[t_1,t_2)$, yields the first dependency tree of the
	sequence: $\TimeDepTree[t]{\lib[a]}=\DepTree{\lib[a_1]}$. Instead, for
	$t'\in[t_2,t_3)$ we get $\TimeDepTree[t']{\lib[a]}=\DepTree{\lib[a_2]}$.
\end{example}

We call the above \emph{time-indexing}: then \TimeDepTree{\lib[a]} in \Cref{ex:TDT:time-indexing} encapsulates the history of dependency trees of \lib[a], accessible via time-indexing.
Orthogonally we can do \emph{library-slicing} by restricting the \TDT to specific dependencies, to retrieve the d-matrix of the corresponding library in the time span $T$.
We denote this explicitly as a restriction, \TimeDepTreeCut{\lib}{\lib'}, which yields the d-matrix $\bflib'=\left\{\bflib[i]'\right\}_{i=1}^n$ bounded temporally by $T=[\tstart,\tend]$ with $\tstart = \min{\trel}(\bflib') \leqslant \max{\trel}(\bflib') = \tend$, where $\min{\trel}(\bflib')\doteq\min_{i=1}^n\big\{\trel\big(\bflib{[i,1]}'\big)\big\}$ is the min time of release among the ($n$ or less) instances of library $\lib'$ present in \TimeDepTree[\tstart]{\lib}, and analogously for $\max{\trel}(\bflib')$ and the max time of release.

\begin{example}
\label{ex:TDT:library-slicing}
	In \TimeDepTree{\lib[a]} from \Cref{fig:TDTs:TDT_example_1},
	the relevant versions of dependency \lib[c] are given by
	$\TimeDepTreeCut{\lib[a]}{\lib[c]} = 
	\big[\lib[c_1]\:\lib[c_1]\:\lib[c_2]\big]$, which contracts to the
	c-chain $\lib[c_1]<\lib[c_2]$.
	Different instances of the same library in dependency trees are
	retrieved equivalently, e.g.\ \TimeDepTreeCut{\lib[a]}{\lib[d]}
	yields the d-matrix depicted in \Cref{fig:d-matrices:d-matrix},
	which can be interpreted as c-chains (\Cref{fig:d-matrices:c-chains})
	or flattened to the set $\big\{\lib[d_i]\big\}_{i=1}^3$
	of all relevant instances of library \lib[d].
\end{example}

\begin{figure}
	\centering
	\begin{minipage}{.6\linewidth}
		\includegraphics[width=\linewidth]{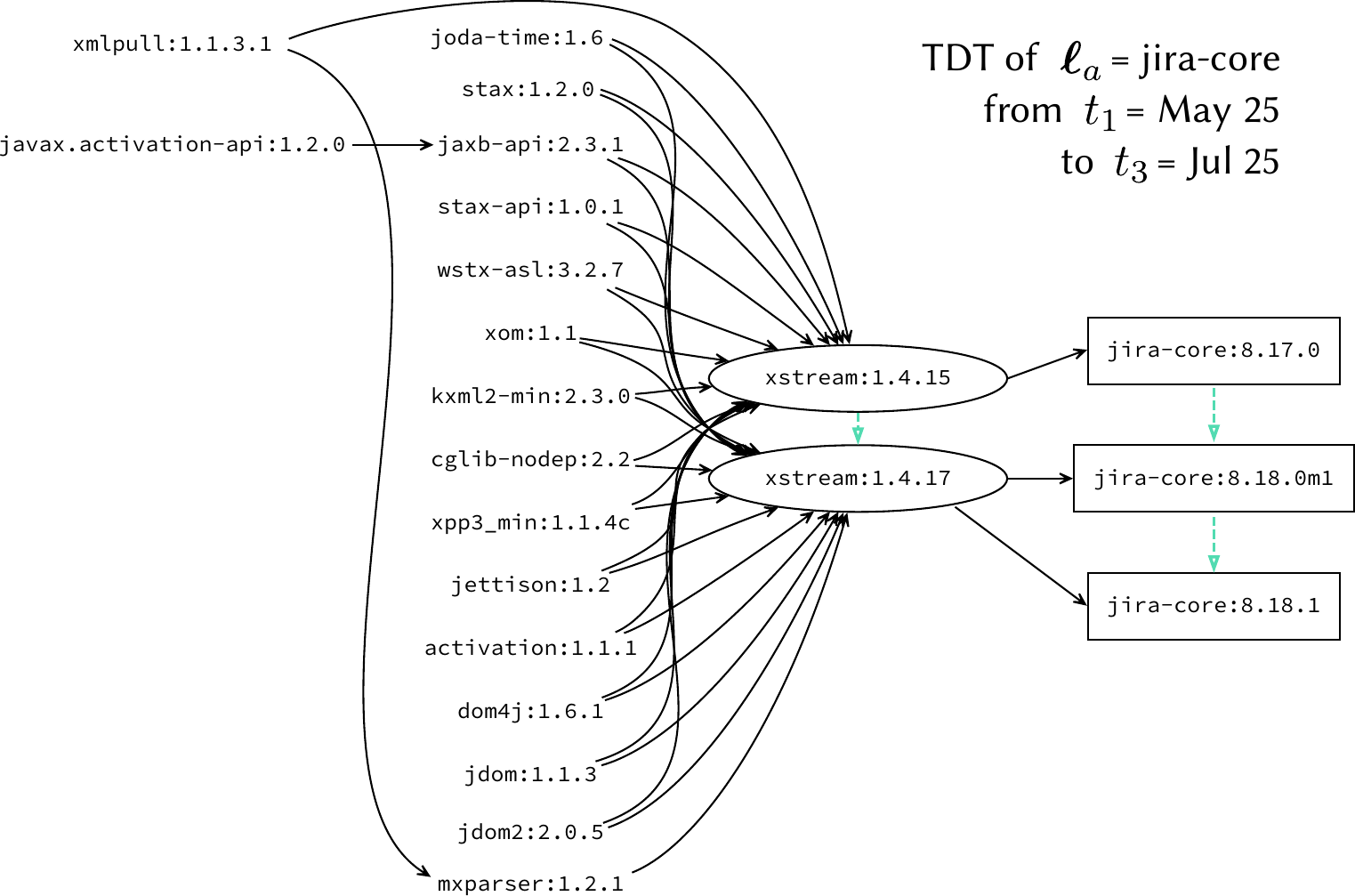}
	\end{minipage}
	\hfill
	\begin{minipage}{.35\linewidth}
		The lack of updates in the dependencies of $\lib[a]=\code{jira-core}$
		is apparent in the \TDT:
		this happens because the two latest versions of \lib[a] both use
		\code{xstream:1.4.17}.

		\begin{hlbox}
		As a consequence, the likelihood of facing a \CVE
		via \code{xstream} will see a peak since version
        \code{8.18.0m1}---see \Cref{fig:fitted_KDE_PDFs}.
		\end{hlbox}

		But all dependencies of \code{xstream:1.4.15}
		are kept for \code{xstream:1.4.17}.

		\begin{hlbox}
		Thus, a \CVE released for any dependency in this \TDT will affect
		\emph{all versions of \lib[a] released between May 25 and Jul 25.}
		\end{hlbox}
	\end{minipage}

	\vspace{-1ex}
	\caption{Portion of the \TDT for the library from
	         \Cref{ex:motivating_example}, for three time points that
			 cover \code{jira-core:\{8.17.0,\,8.18.0m1,\,8.18.1\}}}
	\label{fig:TDT_example}
\end{figure}

\begin{example}
\label{ex:TDT:running_example}
	\Cref{fig:TDTs} illustrates the concepts introduced so far.
	For a real-life example consider the \TDT for the main library from
	\Cref{ex:motivating_example},
	${\lib[a]=\code{com.atlassian.jira:jira-core}}$, and take three-time
	instants: $t_1=\text{May 25}$, $t_2=\text{Jun 25}$, and $t_3=\text{Jul 25}$
	(2021).
	The main library instances are:
	${\lib[a_1]=\code{com.atlassian.jira:jira-core:8.17.0}}$,
	${\lib[a_2]=\code{\ldots 8.18.0m1}}$, and
	${\lib[a_3]=\code{\ldots 8.18.1}}$.
	Restricting the analysis to the dependency of \Cref{ex:motivating_example},
	${\lib[d]=\code{com.thoughtworks.xstream:xstream}}$,
	yields the dependencies
	${\lib[d_1]=\code{xstream:1.4.15}}$, and
	${\lib[d_2]=\lib[{d_3}]=\code{xstream:1.4.17}}$,
	plus all recursive dependencies for a total of 22 library instances,
	counting the own code of \lib[a_k] and \lib[d_k].
	The corresponding \TDT for this analysis is depicted in
	\Cref{fig:TDT_example}
\end{example}

\smallskip\noindent

\smallskip

\section{Estimating the probability of future vulnerabilities}
\label{sec:predict}

The previous \namecref{sec:model} introduced time dependency trees, addressing RQ2.
To address RQ1 (``Can we forecast [vulnerabilities] via metrics that quantify their likelihood in the future?'') this \namecref{sec:predict} shows how \TDTs can be used to estimate the probability that future vulnerabilities affect a library, considering different time points of its history.

We produce forecasts on past events as a means to validate the method---technical details are given here, with an empirical demonstration following in \Cref{sec:appMaven}.
%
\emph{The same approach can be used for future security planning, by proposing (instead of revising) changes in the dependency trees spanned from the \TDT}---we discuss this in the conclusions.

\subsection*{General methodology}
\label{sec:predict:method}

To estimate the probability of vulnerability disclosure in a library, we compare different time points of its \TDT and compute---for each resulting instance---the expected probability of facing a vulnerability in the forthcoming $n$ days.
The input required for this are
\begin{itemize*}[label=,afterlabel=\ ,before=\unskip{:},itemjoin={;},itemjoin*={; and}]
\item	the forecast time horizon in days, $0<n<\infty$
\item	the \TDT of the main library, \TimeDepTree{\lib}
\item	one or more time points to index \TimeDepTree{\lib}
\end{itemize*}.
The output are the probabilities, for \lib at each given time point, of facing a vulnerability within the next $n$ days.
The methodology is summarised next, and detailed in \Crefrange{sec:predict:clusterisation}{sec:predict:estimation}.

\subsubsection*{Overview of methodological steps:}
\begingroup
\setlength{\ULdepth}{.3ex}
\def\INPUT{\emph{\uline{Input}:}\xspace}
\def\OUTPUT{\emph{\uline{Output}:}\xspace}
\def\PROCEDURE{\emph{\uline{Procedure}:}\xspace}

\begin{steps}
\smaller[1]
\item
\label{step:clusterisation}
\textbf{Environment preprocessing (``clustering'')}%
\,\hrulefill\:\Cref{sec:predict:clusterisation}
\\
\INPUT
	\begin{enumerate*}[label=\textsl{(\roman*)},itemjoin={{; }}]
	\item	software development environment, e.g.\ Maven central, PyPI, npm, etc.
	\item	list of measurable properties of the library instances, e.g.\ own size in LoC, cyclomatic complexity, etc.
	\item	bounds for the properties values, e.g.\ ${\text{LoC}\leqslant10k<\text{LoC}}$.
	\end{enumerate*}
\\
\OUTPUT
	clusters $\{C_\iota\}_{\iota\in I}$ of libraries in the environment, as per the requested properties and bounds.
\\
\PROCEDURE
	\begin{enumerate*}[label=\textsl{(\alph*)},itemjoin={{; }}]
	\item	measure the properties of the library instances in scope
	\item	partition the population of libraries, defining one \emph{cluster} per combination of possible property values as divided by the value bounds.
	\end{enumerate*}
\item
\label{step:PDF}
\textbf{Vulnerability data aggregation via PDF fitting}%
\,\hrulefill\:\Cref{sec:predict:PDF}
\\
\INPUT
	\begin{enumerate*}[label=\textsl{(\roman*)},itemjoin={{; }}]
	\item	development environment, as above
	\item	clusters $\{C_\iota\}_{\iota\in I}$ from \cref{step:clusterisation}
	\item	vulnerability records, e.g.\ \acronym{cve}{s} from the \acronym{nvd}.
	\end{enumerate*}
\\
\OUTPUT
	\PDFs $\{f_\iota\}_{\iota\in I}$, one per cluster, for the probability of vulnerability disclosure as a function of time (since library instance release).
\\
\PROCEDURE
	\begin{enumerate*}[label=\textsl{(\alph*)},itemjoin={{; }}]
	\item	map each vulnerability to the (``\emph{evidence}'') library instances that contain them
	\item	compute the time between the release of each evidence library instance and the publication of the vulnerability
	\item	per cluster, fit a \PDF to the dataset of time periods computed.
	\end{enumerate*}
\item
\label{step:TDT}
\textbf{TDT instantiations and conversions to ATs}%
\,\hrulefill\:\Cref{sec:predict:TDT}
\\
\INPUT
	\begin{enumerate*}[label=\textsl{(\roman*)},itemjoin={{; }}]
	\item	\TDT of main library, \TimeDepTree{\lib}
	\item	time points to instantiate the \TDT, $\{t_k\}_{k=1}^m$, with $m\geqslant1$.
	\end{enumerate*}
\\
\OUTPUT
	\ATs $\{\thisAT[k]\}_{k=1}^m$, corresponding to the dependency trees of \lib at time points $\{t_k\}_{k=1}^m$.
\\
\PROCEDURE
	\begin{enumerate*}[label=\textsl{(\alph*)},itemjoin={{; }}]
	\item	index the \TDT at every time point $\{t_k\}_{k=1}^m$, to obtain the corresponding dependency trees $\big\{\TimeDepTree[t_k]{\lib}\big\}_{k=1}^m$
	\item	with the bijection from \Cref{sec:background:AT} generate the corresponding \ATs, i.e.\ one  \thisAT[k] for every $\Treeo_k\doteq\TimeDepTree[t_k]{\lib}$.
	\end{enumerate*}
\item \label{step:estimation}
\textbf{Probabilities estimation}%
\,\hrulefill\:\Cref{sec:predict:estimation}
\\
\INPUT
	\begin{enumerate*}[label=\textsl{(\roman*)},itemjoin={{; }}]
	\item	forecast time horizon in days, $0<n<\infty$
	\item	\PDFs $\{f_\iota\}_{\iota\in I}$ from \cref{step:PDF}
	\item	\ATs $\{\thisAT[k]\}_{k=1}^m$ from \cref{step:TDT}.
	\end{enumerate*}
\\
\OUTPUT
	Estimates $\{\hat{p}_k\}_{k=1}^m$, where $p_k$ is the probability that any vulnerability affects \lib within $n$ days counting from the time point $t_k$.
\\
\PROCEDURE
	Each leaf $x$ of an \AT \thisAT[k] from \cref{step:TDT} stands for a dependency $x$ of \lib at time $t_k$:
	\begin{enumerate*}[label=\textsl{(\alph*)},itemjoin={{; }}]
	\item	identify the cluster to which $x$ belongs, and its corresponding \CDF $F_x$ from \cref{step:PDF}
	\item	displace $F_x$ to the left by the release time of $x$, and integrate for $n$ days
	\item	label with that probability the leaf of \thisAT[k] corresponding to $x$
	\item	propagate all such probabilities in the \AT: the result is the probability estimate $\hat{p}_k$
	\end{enumerate*}.
\end{steps}
\endgroup

\subsection{Environment preprocessing (clustering)}
\label{sec:predict:clusterisation}

Each future-vulnerability estimate is computed from a \PDF, that describes the probability that a \CVE is published for an arbitrary library instance (\lib[x]) as a function of time since its release ($f_x\from\RR_{\geqslant0}\to[0,1]$).
Such \PDFs are empirical, fitted from data on the time between the release of \lib[x] and the publication of a \CVE affecting its own codebase%
\footnote{%
	Statistical correctness is ensured by counting each \CVE \emph{once} for the specific codebase that it affects, viz.\ transitive vulnerabilities are not counted.
}.
So e.g.\ if we have two such \CVEs published for \lib[x] after 30 and 65 days of its release, the data-fitting pool for $f_x$ is $P_x=\{30,65\}$.

But such vulnerabilities are rare events in the space of library instances---see \Cref{sec:background:related_work}---meaning that not only \card{P_x} is generally too low to justify the goodness of fit of $f_x$, but even $P_x=\emptyset$ in the vast majority of cases.

A workaround is to aggregate data in the development environment $\mathcal{E}$ of \lib[x], e.g.\ fit a single \PDF $f_C$ from $P_C$ with ${C\doteq\{\lib[x] \in \mathcal{E} \mid \lib[x]~\text{has a \CVE}\}}$.
However, the distribution of vulnerabilities is not homogeneous in the space of library instances of an environment.
In particular, some \lib[x] have \CVEs disclosed often, even monthly, while others have no \CVE altogether.
A partition $\biguplus_{\iota \in I}C_\iota = C$ must then be found, that
$(i)$ yields sufficient data points in each $P_\iota$ to fit a \PDF $f_\iota$, e.g.\ $\card{C_\iota} > 30$, and
$(ii)$ puts library instances \lib[x] and \lib[y] in different $C_\iota$ if they have different security characteristics.

Approaching this as a standard \ML clusterisation problem is error-prone, due to the absence of ground truth on features---code properties---that are relevant for security vulnerabilities.
For example, interpreting $(ii)$ as the problem \textquote{find the partition that minimises the variance of each $P_\iota$ s.t.\ $(i)$ holds} is incorrect: reducing times variance does not mean that one captured the same security aspects in the corresponding cluster of libraries $C_\iota$---see \Cref{sec:discussion:data} for a discussion.

The theoretically optimal solution for this problem is conjectured to depended on the development environment \emph{even for feature importance}, e.g.\ it could happen that the number of lines of code is relevant to separate clusters of Java or Python libraries, but not Haskell ones.
But even if optimal solutions exist, finding them is out of scope.
Instead, our empirical demonstration of the general method in \Cref{sec:appMaven} proceeds pragmatically by selecting two security-linked properties, that are measurable on our corpus of library instances and allow a partition into four clusters  $\{C_\iota\}_{\iota=1}^4$.

\subsection{Vulnerability data aggregation via PDF fitting}
\label{sec:predict:PDF}

Given a c-chain $\lib[v_j]<\lib[v_{j+1}]<\cdots<\lib[v_i]<\lib[v_{i+1}]\cdots$  we say that library instance \lib[v_i] is \emph{evidence} of some \CVE `\CVEn[1]' when \CVEn[1] is a direct vulnerability of the c-chain (i.e.\ it affects the own code of library \lib), and $v_i$ is the last version of \lib where \CVEn[1] is observed.
Given that \CVEs usually affect several related library instances, evidence libraries are needed to count the data point corresponding to a \CVE only once per c-chain.

While any library in the c-chain that is vulnerable to a \CVE serves as evidence, e.g.\ the first one, responsible disclosure makes it common to \emph{publish \CVEs based on the latest code that they affect}---because a patch is first released and the \CVE is published afterwards.
More importantly, when a \CVE \CVEn[1] is found in \lib[v_i], the classical conservative approach is to deem \emph{all previous libraries \lib[v_j] with $j<i$ also vulnerable to \CVEn[1]}---see e.g.\ CVE-2021-39139 in \Cref{ex:motivating_example}.
Thus, using the latest vulnerable version of the c-chain as evidence ($i$) ensures correct counting and ($ii$) is consistent with current practice.
In turn, when a \CVE \CVEn[1] affects more than one c-chain (e.g.\ 
\href{https://www.cve.org/CVERecord?id=CVE-2010-2076}{CVE-2010-2076},
\href{https://www.cve.org/CVERecord?id=CVE-2021-22569}{CVE-2021-22569}, and
\href{https://www.cve.org/CVERecord?id=CVE-2021-33037}{CVE-2021-33037})
then \CVEn[1] is counted as a data point once for each corresponding library evidence.

Let \emph{grace period} denote the number of days between the release of a library instance \lib[v_i], and the disclosure of a \CVE for which \lib[v_i] is evidence.
The \PDF $f_\iota$ of cluster $C_\iota$ is fitted on the grace periods of the library instances in $C_\iota$.
Since this is empirical data we resort to kernel-density estimation \cite{Ros56,Par62}.
Thus, the resulting cumulative density function (\CDF) for cluster $C_\iota$, $F_\iota\from\RR_{\geqslant 0}\to[0,1]$, measures the probability of disclosure of a \CVE up to time $t\geqslant0$ counting from the time of release of a library instance $\lib[v_i]\in C_\iota$.
In other words, $S_\iota(t) \doteq 1-F_\iota(t)$ is the survival function of library instances $\lib[v_i]\in C_\iota$ for the event ``disclosure of a \CVE affecting the own code of \lib[v_i] at time $t$ after the release of \lib[v_i]''.


\subsection{\texorpdfstring{\TDTs}{TDTs} instantiation and conversions to \texorpdfstring{\ATs}{ATs}}
\label{sec:predict:TDT}

To estimate the probability of future vulnerabilities of \lib, we start from its \TDT and index it in the time point(s) requested by the user to retrieve the corresponding dependency tree.
Let $\Treeo_k\doteq\DepTree[t_k]{\lib}$ be one such instantiated dependency tree: by \Cref{def:TDT} it contains all library instances that are dependencies of \lib at time $t_k$.
Applying to $\Treeo_k$ the transformation from \Cref{sec:background:AT} we obtain the corresponding attack tree \thisAT[k], which contains one leaf node---called basic attack step or \BAS---for every library instance occurring in $\Treeo_k$, including the own codebase of \lib at $t_k$.

\begin{figure}
	\centering
	\includegraphics[width=.95\linewidth]{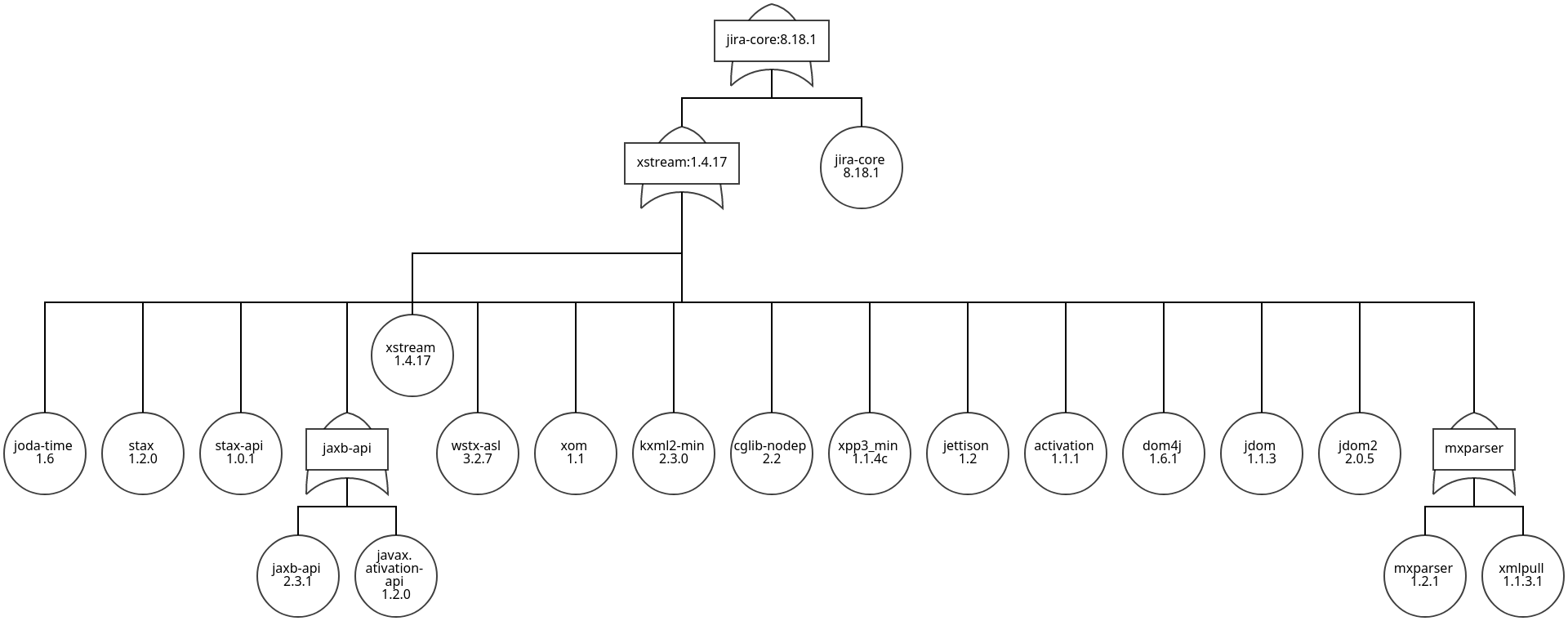}
	\caption{\thisAT[3] = \AT corresponding to time-index $t_3$
	         in the \TDT from \Cref{fig:TDT_example}, i.e.\ the
	         dependency tree of \code{jira-core:8.18.1} ($\Treeo_3$).}
	\label{fig:AT_example}
\end{figure}

\begin{example}
\label{ex:predict:AT}
	Let us show this on the \TDT from \Cref{fig:TDT_example}, which
	studies \code{xstream.com.thoughtworks:xstream} as dependency
	of the main library \code{com.atlassian.jira:jira-core} between
	$t_1=\text{May 15}$ and $t_3=\text{Jul 25}$, 2021.
	At $t_3$, the root of $\Treeo_3$ 
	is the library instance \code{jira-core:8.18.1}.
	It has dependencies, so it is a non-leaf node that is mapped
	to a \BAS leaf and an \OR gate---see \Cref{sec:background:AT}---that
	are shown at the top of \thisAT[3] in \Cref{fig:AT_example}.
	The \BAS is the circle labelled \code{jira-core:8.18.1},
	that stands for the own code base of that library instance.
	The \OR gate has two children: this \BAS and \code{xstream:1.4.17},
	meaning that a vulnerability affecting \code{jira-core:8.18.1} can come
	from its own codebase, or from the direct dependency \code{xstream:1.4.17}.
	But \code{xstream:1.4.17} is also a non-leaf node in $\Treeo_3$, and thus
	represented by its own \OR gate and \BAS---see \Cref{fig:AT_example}.
	This recursion in $\Treeo_3$ ends when the leaf nodes are reached,
	i.e.\ library instances without dependencies such as \code{xmlpull:1.1.3.1}.
	Those are the base cases of the recursion, each mapped to a \BAS
	showing at the bottom of \thisAT[3] in \Cref{fig:AT_example}.
 \end{example}

\subsection{Probabilities estimation}
\label{sec:predict:estimation}

For time point $t_k$ let \thisAT[k] be the \AT corresponding to the dependency tree of \lib at time $t_k$.
Each leaf in \thisAT[k], say $\BAS^x$, stands for a dependency \lib[x] of \lib at time $t_k$.
But library instance \lib[x] was released before $t_k$, so to use the \PDFs fitted during \cref{step:PDF}, whose origin is the release date of the library, we must first displace the \PDF origin by the time elapsed since such release.
For that we measure the time $\Delta^x_k \doteq t_k - \trel(x) > 0$ elapsed between the release of dependency \lib[x] and $t_k$.
Then we compute $\hat{p}^x_k\doteq F_\iota\big(\Delta^x_k+n\big)-F_\iota\big(\Delta^x_k\big)$, where $F_\iota$ is the \CDF of the cluster $C_\iota$ to which dependency \lib[x] belongs.
That is a probability value, that corresponds to the likelihood that a \CVE is released for $\lib[x]\in C_\iota$ within $n$ days since $t_k$.
We decorate $\BAS^x$ in \thisAT[k] with $\hat{p}^x_k$, and move to the next leaf.
Once this has been done for all leaves, these discrete probabilities are propagated in the \AT via standard algorithms: the result is the probability estimate $\hat{p}_k$.

\begin{example}
\label{ex:methodology:estimation}
	Let us estimate the probability of facing a vulnerability in the
	running example, for $t_3=\text{Jul 25}$, 2021 (viz.\ \thisAT[3] from
	\Cref{fig:AT_example}), and for a time horizon of $n=\text{45 days}$.
	The transitive dependency $\lib[x]=\code{xmlpull:1.1.3.1}$---%
	bot\-tom-\-right corner in \Cref{fig:AT_example}---has 1768 LoC
	and is web-oriented, so its \PDF is that of the cluster
	$C_\iota=\text{Small/Medium} \cap \code{Remote\:network}$
	that we introduce next in \Cref{sec:appMaven}.
	We have the corresponding \CDF from \cref{step:clusterisation,step:PDF},
	$F_\iota$, but its origin is located at the release date of this library
	instance: $\trel(\lib[x])=\text{Jun 16, 2003}$, which is quite far back
	in the past.
	The desired probability---covering $n$ days since $t_3$---lies to the right
	of that origin, namely $\Delta_3^x=t_3-\trel(\lib[x])$ days to the right.
	Therefore, we estimate the desired probability for \lib[x] as
	$\hat{p}^x_3
		= F_\iota\big(\Delta^x_3+n\big)-F_\iota\big(\Delta^x_3\big)
		\approx 7.80\cdot10^{-4}$,
	and decorate the corresponding \BAS in \thisAT[3] with that value.
	Repeating this procedure for dependency $\lib[y]=\code{mxparser:1.2.1}$
	yields a much larger estimate, $\hat{p}^y_3=8.36\cdot10^{-2}$,
	mainly because its release date $\trel(\lib[y])=\text{Mar 11, 2021}$,
	is much closer to $t_3$, viz.\ this is a relatively new library instance
	for which a \CVE release is still quite likely in principle.
	Once all probabilities have been estimated for the \BAS{es} in \thisAT[3],
	we propagate the total probability through the \AT \cite{LBS23},
	which yields the final estimate $\hat{p}_3\approx0.158$
	of facing a vulnerability in $n=45$ days since $t_3=\text{Jul 25}$.
	In \Cref{sec:appMaven:TDT} we show that for $t_2=\text{Jun 25}$ one
	gets $\hat{p}_2\approx0.096$, which is coherent with the Maven
	data processed by our methods and, much more importantly,
	is able to foretell the situation found in the motivating example
	from \Cref{fig:motivating_example}.
\end{example}


\section{Application to the Maven Ecosystem}
\label{sec:appMaven}

This \namecref{sec:appMaven} reports the outcomes obtained by applying the methodology from \Cref{sec:predict} to 1255 \CVEs and 768 \FOSS libraries from the Java/Maven ecosystem.
We make our results available---including the execution scripts and partial results for the for \namecrefs{ex:methodology:estimation} in \Cref{sec:predict:method}---as FOSS via a software-reproduction package~\cite{BPM24}.

\paragraph{Computational efficiency}
\Cref{step:clusterisation,step:PDF} can be done for a relevant subset of the environment, and reused in several studies after executing them once.
For instance, this empirical demonstration works on Maven libraries with severe vulnerabilities---viz.\ $\CVSS\geqslant7$---which were split into four clusters.
In contrast, while \cref{step:TDT,step:estimation} have lower runtime and memory footprint, they are recomputed for each new user input.
By way of example, using a standard i7-CPU laptop with 32~GB RAM, our prototypical Python implementation takes a few minutes to perform all these \namecrefs{step:TDT} in our corpus.
Execution of \cref{step:TDT,step:estimation} takes a few seconds---see artifact below.
An efficiency-driven implementation, e.g.\ using the \BDD algorithms from \cite{LBS23}, can scale these results to thousands of dependencies.

\paragraph{Data and Artifact Availability}
We provide public access to all files and scripts generated and used in this work, to make them available for reproduction and reuse of our results, in \cite{BPM24} and in this link:
\href{https://zenodo.org/records/10895796?token=eyJhbGciOiJIUzUxMiIsImlhdCI6MTcxMTc0OTM5MiwiZXhwIjoxNzIzNTkzNTk5fQ.eyJpZCI6Ijg5MmJiMzA5LTI5MzMtNGI0My05ZWFkLWUyOWYzNmZiOGFkYSIsImRhdGEiOnt9LCJyYW5kb20iOiI4ZDcxMDY4ZGRmYzkwYWFlODQ4OGM1MjA1M2Q3ZjkxOCJ9.mzvKRPFIoA7SyAF9QkIB4kRRxJyd1iiMri6FpJCCbVV8R0o0EazN5TTqAJql0NpxkwIINlvTyosgN6CXnWlfMg}{\smaller[.5]\color{blue}$\mathtt{https{:}//zenodo.org/records/10895796}$}.
\!\footnote{%
    \bfred{NOTE to reviewers:}
    this is now shared under restricted access for reviewing purposes only: it will be made public upon acceptance of the manuscript.
    A private \acronym{url} to access it can be found in \cite{BPM24}, or clicking the link above---\emph{the URL} \url{https://zenodo.org/records/10895796} \emph{will not grant permission.}
}

\subsection{Environment preprocessing (clustering)}
\label{sec:appMaven:clusterisation}

\begin{figure}
	\centering
	\begin{tikzpicture}[anchor=north east]
		\node[anchor=north west] at (0,0)
			{\includegraphics[width=.99\linewidth]{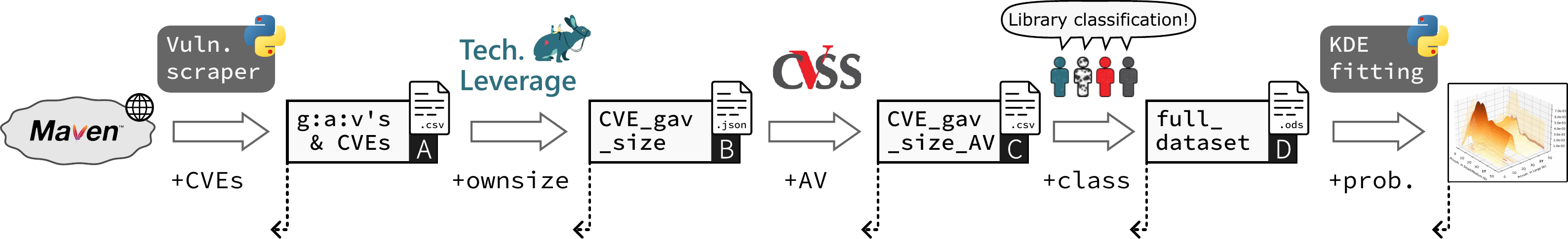}};
		\node[overlay] at ( 2.70,-2.03) {\cite{KGO+18,PPP+22}};
		\node[overlay] at ( 5.61,-2.03) {\cite{MP21b}};
		\node[overlay] at ( 8.36,-2.03) {\cite{JRS+23}};
		\node[overlay] at (10.92,-2.06) {\emph{new!}};
		\node[overlay] at (13.79,-2.06) {\emph{new!}};
	\end{tikzpicture}
	\caption{Data-selection and enrichment process:
	         from Maven libraries to $n$-dimensional vulnerability \PDFs}
	\label{fig:toolchain}
\end{figure}
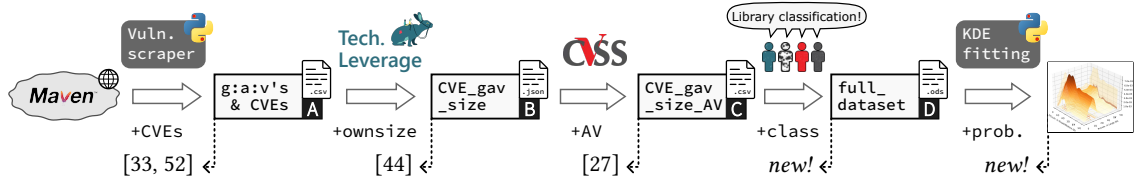

We chose Java/Maven as our target ecosystem.
We selected libraries for which the source code (of their library instances) is attainable via the Maven Central repository, so own codebase size is a usable metric for clusterisation.
Another library-selection criterion was the existence of a \CVE affecting it---because the goal of our experiment was to fit \PDFs to the grace period between the release of a library instance, and the publication of a \CVE for which it is evidence.
We further enrich the data retrieved, through structured processes, as an experimental approximation towards the optimal solution conjectured in \Cref{sec:predict:clusterisation}.
This data gathering, enrichment, and clusterisation, was implemented as follows.

\subsubsection*{Selection and enrichment of Java/Maven libraries}
~~(enumeration matches the data files in \Cref{fig:toolchain})
\begin{enumerate}[label=\raisebox{1pt}{\relscale{.8}\colorbox{black!88}{\parbox{.4em}{\centering\color{white}\sffamily\!\Alph*\!}}},leftmargin=1.8em]
\item	Initial filter via \CVE:
	\begin{itemize}
	\item	we selected \CVE entries, from the Snyk database \cite{SnykDB},
			that affect Java code and have high- or critical-severity;
	\item	this matched 1255 entries that affect 36040 \gav's in Maven,
			corresponding to 768 unique Java libraries (\ga's).
	\end{itemize}
\item	Evidence libraries with available codebase:
	\begin{itemize}
	\item	from the 36k library instances, we cross-checked the affected
			version range of each \CVE entry
			and the \gav data entries in Maven Central, to select those
			that are evidence of a \CVE---see \Cref{sec:predict:PDF}%
			\footnote{%
				While 17\% of the \CVEs 
				affect multiple c-chains, thus spawning multiple evidence,
				13\% of the \gav's 
				are affected by multiple (as many as 11) \CVEs.};
	\item	the selected library instances were checked for source-code
			availability, and those available in Maven had their own codebase
			size measured via the TechLeverage \acronym{api} \cite{MP21b};
	\item	the results were 394 evidence library instances (with an own-size),
			corresponding to 256 unique Java libraries.
	\end{itemize}
\item	Network exploitability of \CVE:
	\begin{itemize}
	\item	those library instances were enriched with their own-size,
			but still lacked data on their web-orientation,
			which is one of the dimensions that will be used in clusterisation;
			as a necessary first step in that direction, we queried the
			\NVD database to determine whether the \acronym{av} metric
			of the corresponding \CVEs is \code{Network} or not.
	\end{itemize}
\item	Web-orientation of evidence library instances:
	\begin{itemize}
	\item	the final step is to classify each library as web-oriented
			(codename \code{Remote\:network}) or not (\code{Local});
	\item	this was performed, for the 256 unique Java libraries,
			via the structured process from \cite{PFMB24}.
	\end{itemize}
\end{enumerate}

\subsubsection*{Clusterisation of the selected libraries}
\begin{figure}
	\centering
	\includegraphics[width=.5\linewidth]{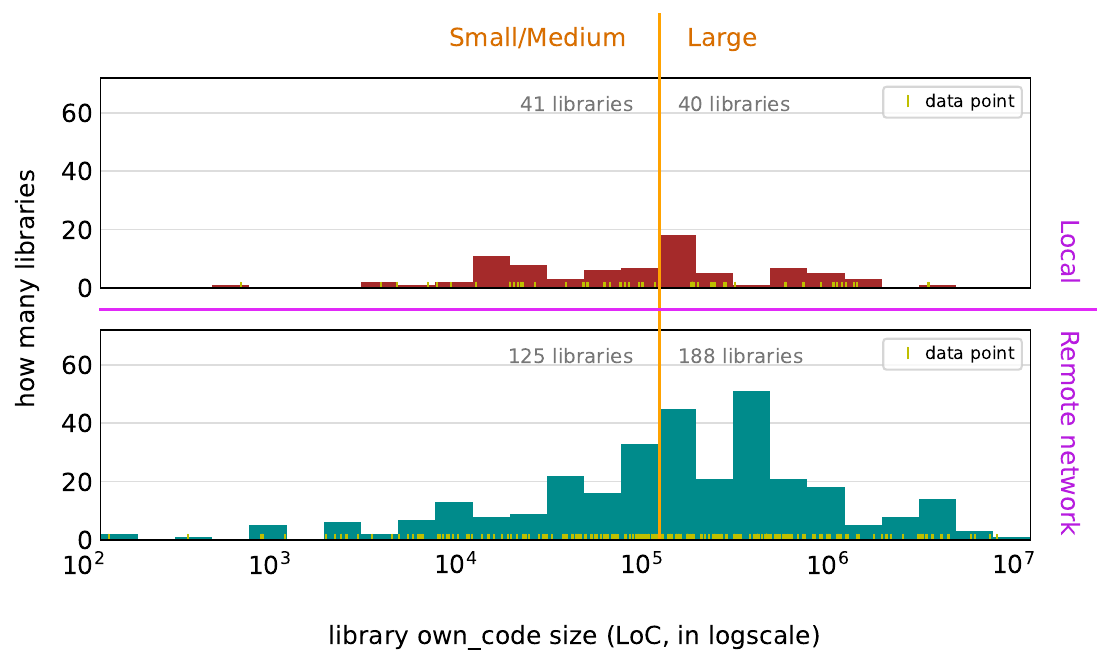}
	\hfill
	\begin{minipage}[b]{.47\linewidth}
	Library instances were split in four quadrants $\{C_\iota\}_{\iota=1}^4$
	according to their own codebase size ($\leqslant100k$ LoC) and
	web-orientation.
	\code{Remote\:network} libraries are more abundant than \code{Local}
	ones due to our selection of \gav{s} with high- or critical-severity \CVE{s}.
	In numbers:
	\begin{itemize}
	\item	Local vs.\ Remote network: 81 vs.\ 313
	\item	Small/Medium vs.\ Large: 166 vs.\ 228
	\item	Local $\cap$ Small/Medium = 41
	\item	Local $\cap$ Large = 40
	\item	Remote network $\cap$ Small/Medium = 125
	\item	Remote network $\cap$ Large = 188
	\end{itemize}
\smallskip
	\end{minipage}

\vspace{-1ex}
	\caption{Histogram of libraries used for clustering:
	         own code size vs.\ web-orientation}
	\label{fig:histogram_clusters}
\end{figure}

Data selection and enrichment resulted in 394 Java library instances, that are evidence to at least one high- or critical-severity \CVE, and whose own-size and web-orientation is known.
These two properties allowed us to divide the corpus into four quadrants such that, for a given library instance \lib[i]:
\begin{enumerate*}[label=\textsl{\bfseries(\alph*)},itemjoin={{; }},itemjoin*={{; and }}]
\item \label{split:size}
\textbf{its \emph{own-size}} is classified as either Small/Medium ($\ownsize[i]\leqslant100k$) or Large%
%
\item \label{split:web}
\textbf{its \emph{web-orientation}} is either \code{Remote\:network} (i.e.\ will likely be directly exposed to the Internet) or \code{Local}.
\end{enumerate*}
Each \lib[i] thus falls in one of the four \emph{clusters} $\{C_\iota\}_{\iota=1}^4$ that result from this cross-product between own-size and web-orientation, as shown in \Cref{fig:histogram_clusters}.

For \ref{split:size}, while literature usually correlates code size to code security \cite{CKDR21,GOP21,SAC21}, different languages have different thresholds to classify libraries by size.
We follow \citet{MP21b} who, studying security risks of vulnerabilities in Java/Maven dependencies, found evidence of qualitative differences among libraries larger or smaller than $100k$ LoC.
In detail, technical-leverage shows how this threshold sets apart libraries \textquote{that mostly prefer to adopt third-party dependencies}, viz.\ $\ownsize[i]\leqslant100k$, from those that \textquote[MP21b]{tend to increase the size of their libraries}.
This is important: we study vulnerabilities rooted in the (own) source code of a library, and must consider qualitative changes in development behaviour that affect that code.
The $100k$ LoC threshold is a language-specific example to distinguish between Small/Medium libraries---where vulnerabilities will most likely come from dependencies---from Large ones---where vulnerabilities become increasingly likely to originate from the own code.

For \ref{split:web}, the expectation is that web-oriented libraries receive more attention and, orthogonally, large libraries increase the chances of an offending line and hence of vulnerabilities.
Small web-oriented libraries are expected to second these, and non-web oriented libraries---both large and small---should be the ones were vulnerabilities are the most scarce, appearing quite late after library release (``if you look, you find, but one seldom looks'').
This classification employed a mapping of PyPI \textsl{Topic classifiers} to Maven \gav{s}, that identifies libraries' main functional purpose through an expert-driven structured procedure \cite{PFMB24}.
The resulting categories
\begin{itemize*}[label=,afterlabel=\ ,before=\unskip{},itemjoin={,},itemjoin*={, and}]
\item	\emph{Communications}
\item	\emph{Database}
\item	\emph{System}
\item	\emph{Security}
\item	\emph{Utilities}
\item	\emph{Internet}
\end{itemize*}
were then classified as \code{Remote\;network} (i.e.\ web-oriented), while the rest are \code{Local}  \cite{PFMB24}.

\subsection{Vulnerability data aggregation via \texorpdfstring{\PDF}{PDF} fitting}
\label{sec:appMaven:PDF}

The output of the previous step were four clusters $\{C_\iota\}_{\iota=1}^4$, each lumping together evidence library instances in the same own-size and web-orientation class.
This data was enriched with the grace periods, i.e.\ the number of days that passed between the release of those library instances, and the publication of a \CVE for which they are evidence.

Thus, for each cluster, we obtain a pool $P_\iota$ of natural numbers equal to the grace periods of the $\lib[i]\in C_\iota$.
A Kernel Density Estimate $K_\iota$ was fit from each $P_\iota$, since \acronym{kde}{s} approximate empirical data for which there are no prior assumptions other than its probabilistic nature \cite{Ros56,Par62}.
We implemented a Python script that performs these operations, using the \code{KDEUnivariate} class from the \code{nonparamtric} module of the \code{statsmodels} library \cite{SP10}.

\begin{figure}
	\centering
	\includegraphics[width=.63\linewidth]{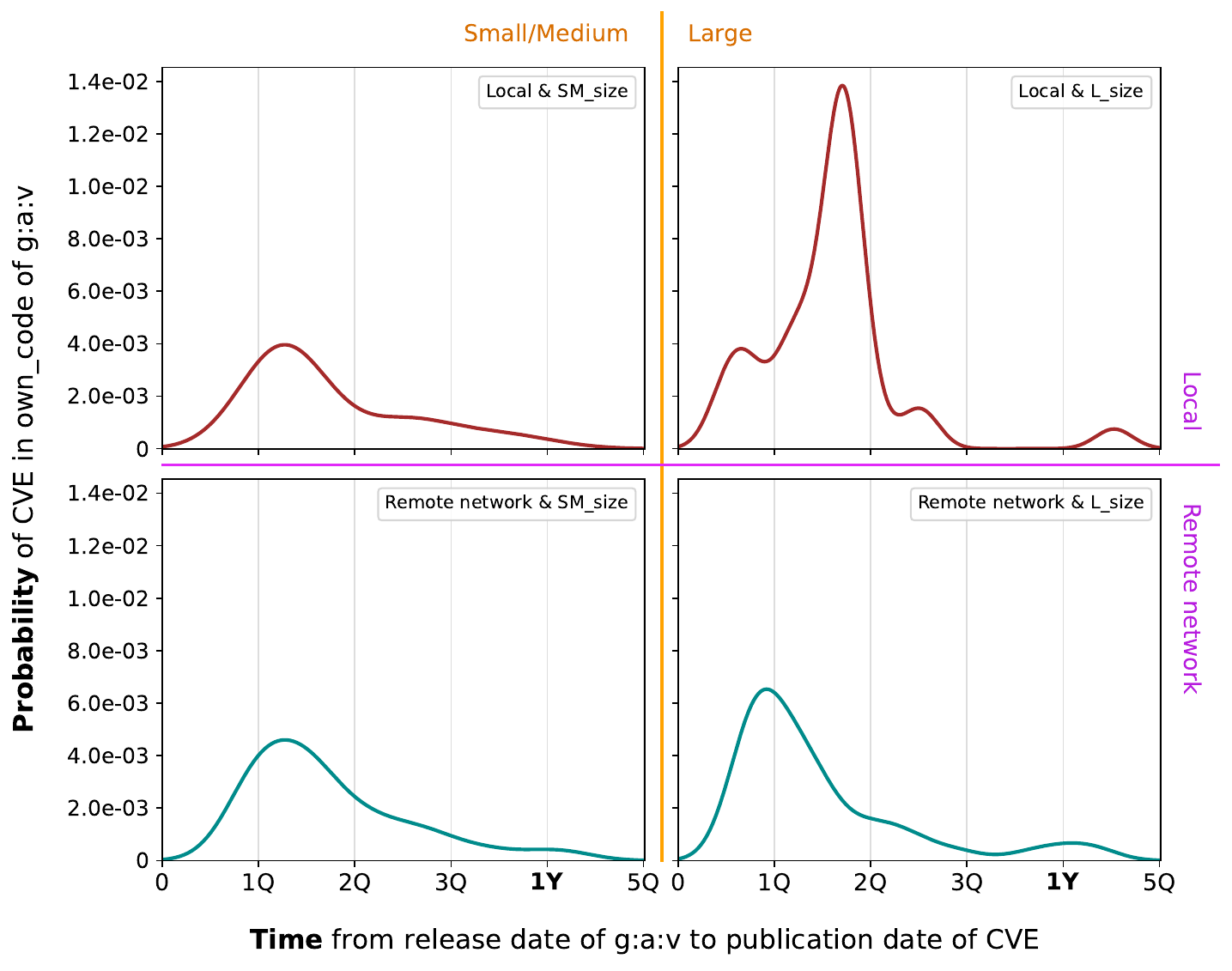}
	\hspace{.5em}\vspace{-1ex}
	\begin{minipage}[b]{.33\linewidth}
The $x$ axis is the time between the release of a \gav, and the publication of a \CVE for which that \gav is evidence---see \Cref{sec:predict:PDF}.
Libraries are divided by web-orientation (\code{Local} vs.\ \code{Remote\:network}) and own-code size (Small/Medium, $\leqslant100k$ LoC, vs.\ Large):
\begin{itemize}[leftmargin=1em,topsep=1.5ex,parsep=1.5ex]
\item
The left skew for \code{Remote\:network} Large libraries suggests higher attention by the security community on web-exposed libraries, as expected.
\item
Small/Medium libraries show similar \CVE publication times regardless of web-orientation, which could be due to the fewer LoC that need revision.
\\[2ex]
\end{itemize}
	\end{minipage}
	\caption{\PDF for each cluster of libraries (class $\times$ size)
	         fitted using \acronym{kde}}
	\label{fig:fitted_KDE_PDFs}
\end{figure}

From $K_\iota$ one can then obtain an approximated probability density function $f_\iota$, and its corresponding \CDF $F_\iota$, that describes the probability of having a \CVE publication as a function of time, from the release date of any library instance belonging to the cluster $C_\iota$ from which $K_\iota$ was fitted.
The result of this process is presented in \Cref{fig:fitted_KDE_PDFs}.
This information can be extrapolated to further libraries, say $\lib[x]\notin C\iota$, as long as \lib[x] can be identified as belonging to cluster $C_\iota$ by virtue of its own-size and web-orientation.

\subsection{\texorpdfstring{\TDTs}{TDTs} instantiation and probability estimation}
\label{sec:appMaven:TDT}

\Cref{ex:methodology:estimation} illustrated how one can use this machinery to estimate the probability of facing a vulnerability in the running example.
This was done for $t_3=\text{Jul 25}$, 2021, and for a time horizon of $n=\text{45 days}$.
The same approach can be followed e.g.\ for $t_2=\text{Jun 25}$, in which case the root of the corresponding dependency tree is \code{jira-core:8.18.0m1}.
Looking at the \TDT from \Cref{fig:TDT_example} one can see that the only library instance that changes between these two time points is that root.
However, another changing factor w.r.t.\ the analysis for $t_3$ in \Cref{ex:methodology:estimation} is the time elapsed since the release of all the library instances---and \Cref{fig:fitted_KDE_PDFs} shows how this time difference affects the probability of facing a \CVE.

Let us study e.g.\ the individual dependency $\lib[y]=\code{mxparser:1.2.1}$ (bottom-right in \Cref{fig:AT_example}).
With 12561 LoC it is classified as Small/Medium sized, \code{Remote\:network}---it is a fork of the XmlPull parsing engine \code{xpp3}, see \cite{PFMB24}---and Maven indicates that it was released on Mar 11, 2021.
Therefore, it's probability estimate since $t_2$ uses $\Delta_2^y=t_2-\trel(\lib[y])=106~\text{days}$, resulting in
$\hat{p}^y_2
		= F_\iota\big(\Delta^y_2+n\big)-F_\iota\big(\Delta^y_2\big)
		\approx 5.52\cdot10^{-2}$.
This is almost half of $\hat{p}^y_3\approx 8.36\cdot10^{-2}$, which the same library instance yields for $t_3$ where $\Delta_3^y=t_3-\trel(\lib[y])=136~\text{days}$.

In contrast, when the release of a dependency library instance was over 15 months ago, computations via our empirical \PDFs for this Java dataset (\Cref{fig:fitted_KDE_PDFs}) are less informative.
This happens for example for \code{jaxb-api:2.3.1}, which was released on Dec 9, 2008, and which here receives the same residual probability $5.79\cdot10^{-4}$ for $t_2$ and $t_3$.

However, it does not take many cases like $\lib[y]=\code{mxparser:1.2.1}$ to increase the probability of facing a \CVE in the full dependency tree.
In particular, $\lib[z]=\code{xstream:1.4.17}$ is a Large (151629 LoC) \code{Remote\:network} library released on May 14, 2021, whose estimate $\hat{p}^z_2\approx0.032$ at $t_2$ increases to $\hat{p}^z_3\approx0.070$ at $t_3$.
Plugging in all values in the \ATs \thisAT[2] and \thisAT[3] of the main library, we estimate a total probability $\hat{p}_2\approx0.096$ of facing a \CVE in 45 days since $t_2=\text{Jun 25}$, which is 40\% lower than the value $\hat{p}_3\approx0.158$ that we obtain for $t_3=\text{Jul 25}$.
The \ATs also reveal that this probability increase (from $t_2$ to $t_3$) in the entire project comes mainly from dependencies $\lib[y]$ and $\lib[z]$.

In other words, our estimators detect a rise---for a time window of 45 days since the end of June---of the probability of a vulnerability affecting ${\lib[a]=\code{jira-core}}$.
Moreover, the analysis shows that this rise comes mainly from its dependencies \code{mxparser:1.2.1} and \code{xstream:1.4.17}---so updating them would ``reset their \PDFs'', reducing the probability of facing a \CVE release in the coming period.
This aligns well with the hindsight of \Cref{ex:motivating_example}, and could have been produced before Aug 2021.
Thus, developers of \code{jira-core} could have used this estimated quantity as motivation to \emph{plan the adoption of any update of library \code{xstream}}, which would have kept the project safe from CVE-2021-39139.

\begin{hlbox}
These time-rolling probability estimates are one example of the type of metric that \TDT analysis can produce, in benefit of decision-making for security planning in software development.
\end{hlbox}


\section{Full Analytical Analysis}
\label{sec:analyanal}

\AT-conversion is best for scalability, and is expected to help planning in production cycles that have predetermined durations.
However, when the dependencies involved are very few, it might be feasible and insightful to perform an analytical study of the probability spaces involved.

Consider for instance a situation in which our interest lies in a single dependency \lib[B] of our project \lib[A], e.g.\ because \lib[B] manipulates sensitive data and is thus posing the greatest security threat.
Say further that \lib[A] has a codebase with less than 100k LoC, so it belongs to the Small/Medium class, while \lib[B] is Large, but both are web-oriented (\code{Remote\:network}).

\begin{figure}
	\centering
	\begin{subfigure}[b]{.36\linewidth}
	\centering

	\includegraphics[width=\linewidth]{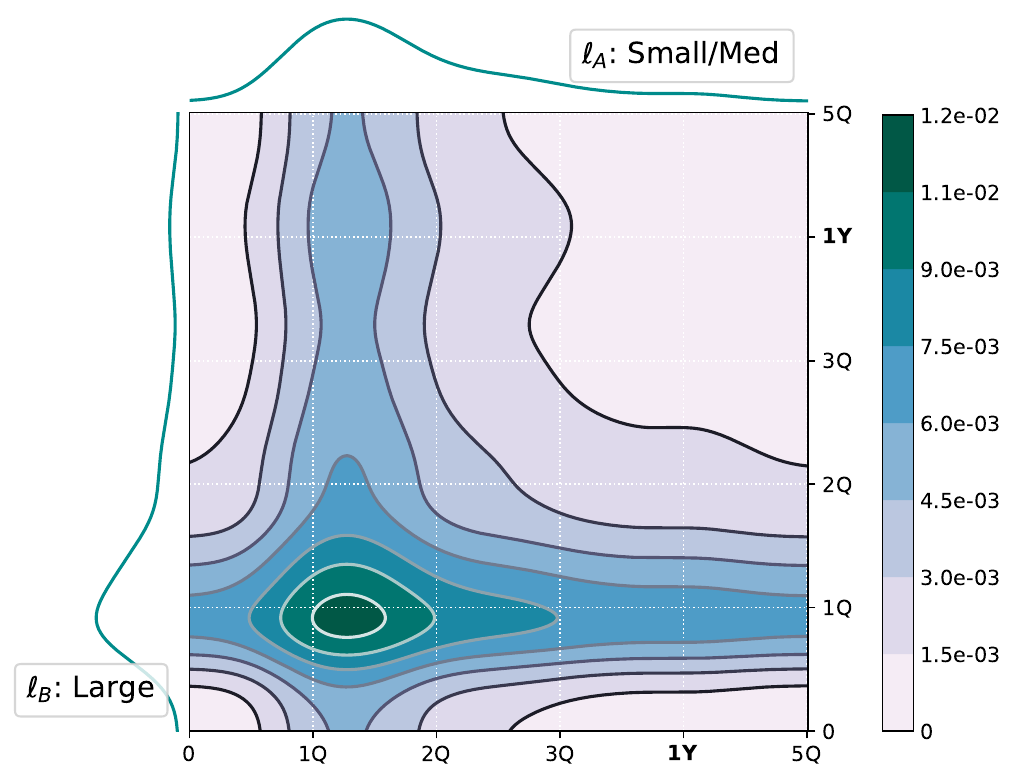}
	\caption{Remote network libraries}
	\label{fig:joint_PDFs:Remote_network}
	\end{subfigure}
	~
	\begin{subfigure}[b]{.36\linewidth}
	\centering
	\includegraphics[width=\linewidth]{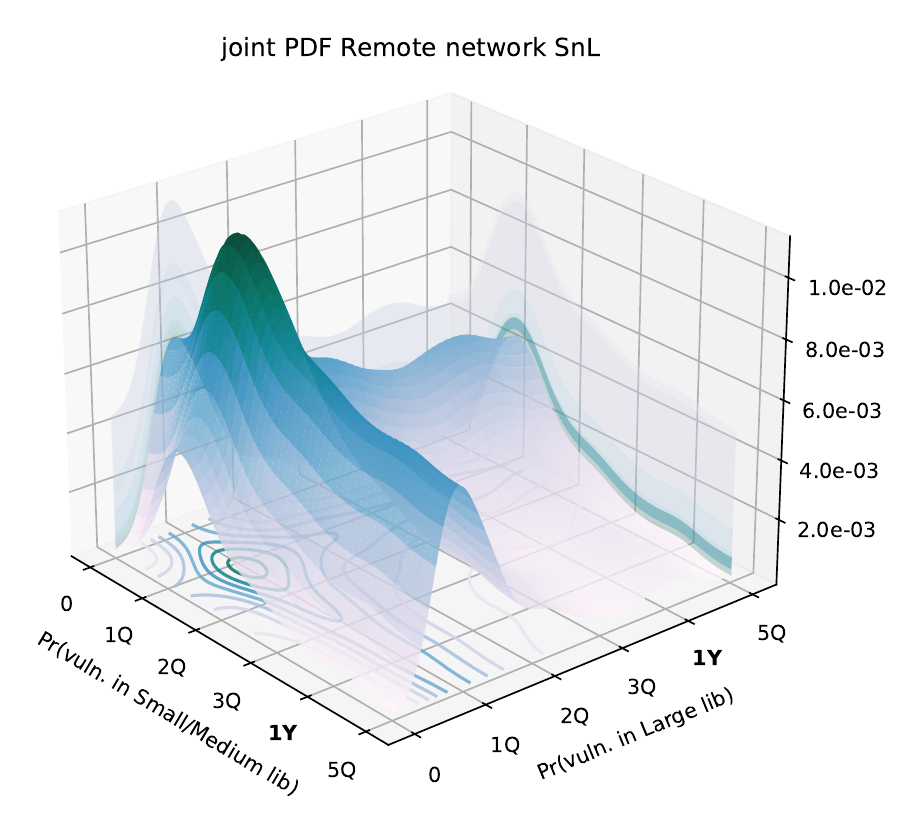}
	\caption{3D projection of the left plane}
	\label{fig:joint_PDFs:Remote_network:3D}
	\end{subfigure}
	\hfill
	\begin{minipage}[b]{.23\linewidth}
	Darker regions indicate higher probability of a \CVE from any
	dependency.
	Remote network libraries are more skewed towards the origin,
	suggesting higher attention by the security community.
	Updating:
	\begin{itemize}[leftmargin=1em,itemsep=.5ex]
	\item	before 1~month keeps all probabilities low;
	\item	is of little use after 6~M.
	\end{itemize}
	\smallskip
	\end{minipage}
	
	\caption{Probability of facing a \CVE from two source codes,
	         one Small/Medium and one Large,
	         applying \Cref{eq:convolution} to \Cref{fig:fitted_KDE_PDFs}}
	\label{fig:joint_PDFs}
\end{figure}

Let $f_A$, $f_B$ be the probability density functions of having a \CVE published for these libraries instances, as fitted by our \cref{step:clusterisation,step:PDF} previously described.
Since a vulnerability discovered in either codebase (ours or theirs) is what we are trying to foresee, the problem can be stated as a race condition between the events ``\CVE released for \lib[A]'' and ``\ldots for \lib[B]''.
Let $F_{A+B}$ denote the \CDF that describes the fastest of those events, i.e.\ $F_{A+B}(t)$ is the probability that at time $t$ we have already observed any of them.
That can be restated as the complement of having survived (i.e.\ not having observed any of them) up to time $t$, which for independent $F_A$ and $F_B$ can be expressed as follows:
\begin{align}
	F_{A+B}(t) = 1-S_A(t)S_B(t) = 1 - \big(1-F_A(t)\big) \big(1-F_B(t)\big).
	\label{eq:convolution}
\end{align}

\begin{figure}
	\centering
	\includegraphics[width=\linewidth]{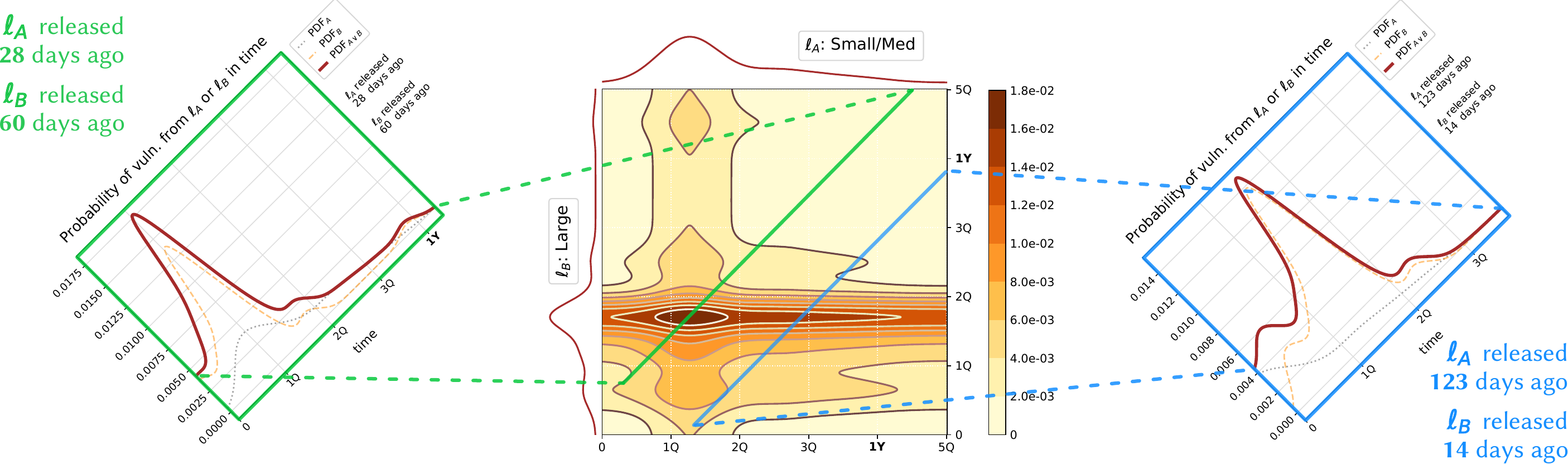}
	\caption{Probabilities of vulnerability from \lib[A] or \lib[B], from
	         class \code{Local}, for different release dates of the libraries}
	\label{fig:joint_PDF:different_release_date}
\end{figure}

This convolution of two \PDFs generates a probability plane, as the ones shown in \cref{fig:joint_PDFs,fig:joint_PDF:different_release_date}, whose $45^\circ$ diagonal cut (viz.\ time passes equally fast for both libraries) is the \PDF of the event ``fastest \CVE released, for \lib[A] or \lib[B]''.
The benefit of this approach with respect to \AT analysis is that it provides the whole \PDF shape, not just discrete data for the $m$ time points $\{t_k\}_{k=1}^m$, and therefore local minima and maxima are apparent.

Taking the main diagonal cut of the plane, that intersects the origin (0,0), represents the case in which \lib[A] and \lib[B] were released on the same day.
One can also displace that diagonal line horizontally by $x_1$ days to the right, to consider a situation in which the library instance plotted on the $x$ axis (here, \lib[A]) was released $x_1$ days earlier.
Similarly, displacing the cut vertically by $y_1$ days upwards represents a release of the other library $y_1$ days earlier.


These analyses are depicted in \Cref{fig:joint_PDF:different_release_date}, which shows probability density curves that instantiate the \PDF convolutions of (\code{Local}) libraries \lib[A] and \lib[B], which were released on different dates in the past:
\begin{itemize}
\item	the green line that cuts the gradient to the left above the main
        diagonal represents the case in which \lib[A] (the Small/Medium
        library) was released 28 days ago, and \lib[B] (the large library)
        was released 60 days ago;
\item	the blue line below to the right represents the case in which
        \lib[A] was released 123 days ago and \lib[B] 14 days ago.
\end{itemize}

Thus, the convoluted probabilities plane is a versatile solution worth considering when the number of codebases involved is low.
While our examples above concern two libraries only, this can be extended to $N$ libraries---at the cost of exponentially-increasing computation power and memory footprint---via the inclusion/exclusion principle as follows:
\begin{align*}
	F_{A_1+\cdots+A_N}(t)
		= \sum_{J\subseteq\{1,\ldots,N\}} (-1)^{\card{J}+1}
		                                 \card{F_{\sum\{A_i \mid i\in J\}}(t)}.
\end{align*}

\section{Discussion}
\label{sec:discussion}

\subsection{Uses of Time Dependency Tree models}
\label{sec:discussion:TDTs}

\Crefrange{sec:predict}{sec:analyanal} show how \TDTs can be used to forecast the disclosure of security vulnerabilities, by producing estimates of the probability of facing them in the future, anywhere in the dependency tree of a project.
However, \TDTs are succinct and general models that allow for a much richer variety of security analyses:
\begin{itemize}
\item
An out-degree count in the nodes---i.e.\ library instances---can \emph{determine the presence of pervasive dependencies}, whose exploitation poses a threat to large portions of a project \emph{across both libraries and time}.
Furthermore, since \TDTs offer a minimal representation this is also computationally efficient.
	\begin{itemize}
	\item	This can be extended to more complex (but still computationally efficient) analyses, such as counting shortest paths to other nodes to cater for transitive dependencies.
	\end{itemize}
\item
\TDTs can be used to \emph{discover single-points-of-failure (\SPOF{s})}, by performing a reverse-reachability analysis across dependency edges.
For example in \Cref{fig:TDTs:TDT_example_2}, a vulnerability affecting library instance $z_{2.1}$ can compromise simultaneously four versions of the main library \lib.
	\begin{itemize}
	\item	This is important because several versions of a library coexist, so exploited \SPOF{s} can cause great disruptions.	For instance in \Cref{fig:TDTs:TDT_example_2}, if some vulnerability is disclosed for $z_{2.1}$ and we were using \lib[1.0] as a dependency, we may consider upgrading to \lib[1.1] or downgrading to \lib[0.9], but none of those actions would fix the issue.
	\end{itemize}
\item
A vulnerability can (and usually does) affect several versions of a library, since the offending lines may remain unchanged from one version to the next.
This is captured by \TDTs via c-chains and d-matrices, whose study can be used to \emph{measure the health of a development ecosystem}.
	\begin{itemize}
	\item	For example, measuring the portions of c-chains---from popular libraries---that were affected by published vulnerabilities, can serve to assess the security risk of developing code in a specific ecosystem.
	\end{itemize}
\end{itemize}

\subsection{Data and white-box models}
\label{sec:discussion:data}

The forecasts produced are a progression of the vulnerability-disclosure probabilities faced by a library \lib in a time window of $n$ days since some date $t$.
For this, our white-box model uses instances of the dependency tree of \lib.
This means that \TDTs can be used to study \FOSS, or by the project owners of \lib---our particular examples here study Java libraries from Maven central.
We use publication of \CVE entries as the ``vulnerability disclosures'' in our model.

Regarding data and as discussed in \Cref{sec:predict:clusterisation}, optimal clustering depends on information that could lie outside the available features.
While this limits the precision of practically-attainable estimators, \emph{more care should be put in avoiding an overfit caused by using too much data.}
\Cref{sec:predict:clusterisation} argues why minimising the variance of data within a cluster is a misguided optimisation approach, that could easily result in overfitting.
Another example would be to convolute the prediction, using probability estimates as part of the data to define clusters: this label creep will make the model overfit, and lose its ability to predict well across projects.
In general, clusters must be defined from independent data sources.

There are, moreover, practical disadvantages in using all available data to develop a prediction model.
Already many of the methods from \Cref{tab:related_work} require expert knowledge (in software development and statistics/ML), hours of machine processing, or even human audits.
This is unavoidable when the model must analyse the entire codebase of a library, or the commit history of years of team development.
In contrast, exploiting the higher-level view provided by the dependency tree yields a good trade-off between fine granularity (\emph{this library instance is dangerous}) and efficient processing (\emph{estimates computed in minutes}).

Automatic procedures can be implemented using the compile tools of the chosen environment, e.g.\ the Maven or Gradle tools for Java libraries.
For instance, the \code{mvn~dependency:tree} command was used in our experimentation to instantiate the \code{xstream} branch of \code{jira-core} at time points $\{t_k\}_{k=1}^3$, for which we constructed the dependency trees.

In-house studies can instead leverage project information available in the company's \acronym{ide}, which is typically more rich since they operate with code that already compiles.
This facilitates the \TDT construction task even further, since the time-version information is typically available in the version-control system used.
Construction of the \AT from the corresponding (instantiated) dependency trees is equivalently approachable.

\subsection{Theoretical extensions}
\label{sec:discussion:highanddry}

On a less technical note, \Cref{sec:background:AT} outlined connections between \TDTs and literature on attack trees, mostly due to the abundant studies on quantitative analysis of the latter, which aligns well with our purpose of estimating the probabilities of potential attacks from \DAG-like attack-oriented models.
We are aware of two further directions in which our work can be reinterpreted, that may yield interesting lines of investigation.

The first is \emph{abstract interpretation theory}---see \citet{Cou21}---where one could set libraries to be the abstractions and library instances their ground terms.
That is, given library \lib spawning a c-chain of library instances $\lib[v_1] < \lib[v_2] < \cdots < \lib[v_{n-1}] < \lib[v_n]$, one can set \lib to be the abstraction of any \lib[v_j], denoted $\abstraction(\lib[v_j])=\lib$, and the library instances \lib[v_j] to be the ground terms of \lib, denoted $\ground(\lib)=\big\{\lib[v_j]\big\}_{j=1}^m$%
\footnote{%
This satisfies the Galois connection since $\abstraction(\ground(\lib))=\lib$, and $\ground(\abstraction(\lib[v_j]))\ni\lib[v_j]$ for all $1\leqslant j\leqslant m$.
}\!.
In a more general sense, abstract interpretation theory could be used to define semantics for refinements or abstractions of \TDTs, viz.\ where specific information is added or abstracted away from the base model, thereby yielding different computation algorithms.

The second possible connection of this work is with \emph{graph transformations} or \emph{rewriting}---see e.g.\ \cite{KNPR18}---whose attributes could be used to formalise restrictions in \TDT evolution.
Some examples are:
\begin{itemize*}[label=\textbullet,itemjoin={{; }}] 
\item	if a node \lib is changed from one dependency tree to the next, all its ancestors must either change (to include the updated version of \lib), or the tree structure must also change (to differentiate those ancestor that keep using the old version of \lib)
\item	time as an attribute allows to express assumptions such as ``dependencies change one at a time, i.e.\ in two successive time steps of the \TDT, at most one leaf version changes''
\item	\CVSS as an attribute may be used to foretell changes in the \TDT, e.g.\ if some node exhibits a critical vulnerability, it should be expected to change (along with its ancestors) in the next step.
\end{itemize*}

\section{Conclusion and future work}
\label{sec:conclu}

We introduced a formal model that yields security-vulnerability forecasts at library level, thus answering RQ1 positively.
These forecasts estimate the probability of facing a \CVE disclosure in a future time window, and consider the own-code of the affected library as well as any of its dependencies.
For that, a white-box model of \emph{time dependency trees} is defined to represent the links between libraries across time (\emph{c-chains} of code evolution) and codebases (library dependencies).
Time dependency trees are coupled with empirical data on past-vulnerability releases, to produce generalisable estimators that cover the research gaps identified in the literature.
This high-level perspective is a lightweight approach that answers RQ2, as we show with an empirical demonstration on 1255 CVEs and 768 Java libraries.

\subsection*{Future directions}
\label{sec:conclu:future}

As introduced in this work, time dependency trees are highly-versatile tools to model the evolution of software in time, and measure its quantitative implications.
While this work details how they can be used for vulnerability forecasting, \Cref{sec:discussion:TDTs} suggests how \TDTs can be employed to detect red-flag library instances---that pose major security risks in a software project---and generate metrics of the security health in a development ecosystem.
On top of that, we foresee some technical extensions that could be easier to exploit for future research:
\begin{itemize}[leftmargin=1.5em,topsep=0pt]
\item	A d-matrix can contain (parallel) c-chains that originate from the
		same codebase, such as Apache Tomcat versions 8.5 and 9.0.
		Our current model treats them as independent codebases---one could
		also collapse them as a single c-chain and measure how does this
		affect the probability predictions in concrete cases.
\item	A replication study of our empirical demonstration on a different
		programming language, or even a different set of Java libraries,
		could serve as indication of how general are the code features
		chosen to define our four clusters.
\item	(Avoiding overfitting and label creep) the dataset of
		features used for clustering could be extended, e.g.\ code update
		frequency and technical-leverage, to study which features produce
		the best forecasts according to past evidence.
\item	Similarly, using different \ML methods to clusterise the dataset
		(e.g.\ principal component analysis, or uniform manifold approximation
		and projection), or to improve the data fitting (e.g.\ via transformers
		or large language models), are directions worth exploring
		\cite{GFQW23,SGL24}.
\end{itemize}

\begin{acks}
This work was partially supported by the European Union under
the \grantsponsor{MSCA}{MSCA}{https://ec.europa.eu/info/research-and-innovation/funding/funding-opportunities/funding-programmes-and-open-calls/horizon-europe/marie-sklodowska-curie-actions_en} grant \grantnum{MSCA}{101067199 \textsl{ProSVED}},
and
\grantsponsor{NextGenerationEU}{NextGenerationEU}{https://next-generation-eu.europa.eu/index_en} projects \grantnum{NextGenerationEU}{D53D23008400006 \textsl{Smartitude}} funded under MUR PRIN 2022
and
\grantnum{NextGenerationEU}{PE00000014 \textsl{SERICS}} funded under MUR PNRR,
and
%
\grantsponsor{HORIZON Europe}{HORIZON Europe}{https://cordis.europa.eu/programme/id/HORIZON.2.3/en} 
\grantnum[https://www.sec4ai4sec-project.eu/]{HORIZON Europe}{101120393 \textsl{Sec4AI4Sec}}.
Views and opinions expressed are however those of the author(s) only and do not necessarily reflect those of the European Union or The European Research Executive Agency.
Neither the European Union nor the granting authority can be held responsible for them.
\end{acks}

\bibliographystyle{ACM-Reference-Format}
\bibliography{paper} 

\appendix

\section{Literature search}
\label{sec:literature_research}

Search string used in \href{https://www.scopus.com}{the Scopus database} to retrieve works relevant to our studies: 

\begin{hlbox}
\ttfamily\smaller
TITLE-ABS-KEY(
	( security OR cybersecurity OR cyber-security )
	AND ( vulnerability OR vulnerabilities )
	AND ( metric OR java OR project )
	AND ( update OR version OR release ) )
	AND PUBYEAR > 2007	
	AND PUBYEAR < 2024
	AND ( LIMIT-TO( SUBJAREA, "COMP" )
		OR LIMIT-TO( SUBJAREA, "ENGI" ) )
	AND ( LIMIT-TO( EXACTKEYWORD, "Java Programming Language" )
		OR LIMIT-TO( EXACTKEYWORD, "Security Vulnerabilities" )
		OR LIMIT-TO( EXACTKEYWORD, "Open Source Software" )
		OR LIMIT-TO( EXACTKEYWORD, "Software Security" )
		OR LIMIT-TO( EXACTKEYWORD, "Computer Software" ) )
\end{hlbox}

This produced 118 results in Scopus, which we added to those listed in the survey by \citet{GS17}, in our search for works that introduce methods or models to discover or predict vulnerabilities in code, including correlation studies between project metrics, and number and severity of vulnerabilities.
The results of this search are discussed in \Cref{sec:background:related_work}, and graphically categorised in \Cref{tab:related_work}.

\end{document}